# Self-healing mechanism of lithium in lithium metal batteries

Junyu Jiao[1*], Genming Lai[1*], Liang Zhao[2,5], Jiaze Lu[3], Qidong Li[2,5], Xianqi Xu[1], Yao Jiang[4], Yan-Bing He[2], Chuying Ouyang[4], Feng Pan[1], Hong Li[3†], and Jiaxin Zheng[1†]

[1]School of Advanced Materials, Peking University, Shenzhen Graduate School; Shenzhen 518055, People's Republic of China.

[2]Shenzhen All-Solid-State Lithium Battery Electrolyte Engineering Research Center, Institute of Materials Research (IMR), Tsinghua Shenzhen International Graduate School, Shenzhen 518055, People's Republic of China.

[3]Beijing Key Laboratory for New Energy Materials and Devices, Institute of Physics; Chinese Academy of Sciences, Beijing 100190, People's Republic of China.

[4]Fujian Science & Technology Innovation Laboratory for Energy Devices of China (21C-LAB), Ningde 352100, People's Republic of China.

[5]School of Materials Science and Engineering, Tsinghua University, Beijing, 100084, People's Republic of China

*These authors contributed equally to this work.

†Corresponding author. Email: zhengjx@pkusz.edu.cn; hli@iphy.ac.cn

**Abstract:** Li metal is an ideal anode material for use in state-of-the-art secondary batteries. However, Li-dendrite growth is a safety concern and results in low coulombic efficiency, which significantly restricts the commercial application of Li secondary batteries. Unfortunately, the Li deposition (growth) mechanism is poorly understood on the atomic scale. Here, we used machine learning to construct a Li potential model with quantum-mechanical computational accuracy. Molecular dynamics simulations in this study with this model revealed two self-healing mechanisms in a large Li-metal system, viz. surface self-healing and bulk self-healing, and identified three Li-dendrite morphologies under different conditions, viz. 'needle',

'mushroom', and 'hemisphere'. Finally, we introduce the concepts of local current density and variance in local current density to supplement the critical current density when evaluating the probability of self-healing.

**Introduction**

Li metal has a low electrochemical potential (3.04 V vs. standard hydrogen electrode) and a specific capacity of ≤3,860 mAh $g^{-1}$, making it an ideal anode material for next-generation secondary batteries (*1, 2*). However, the commercial application of Li secondary batteries is hampered by safety concerns and low coulombic efficiency (*3*): the battery can short circuit if the separator is penetrated by uncontrolled Li-dendrite growth (*4*), while the coulombic efficiency is reduced by the formation of 'dead Li' during cycling (*5*). Several characterisation methods have been used to elucidate the dynamic behaviour of Li-metal deposition (*6, 7*). Although researchers generally agree on the atomic-scale behaviour of Li-dendrite growth, most experimental studies have only described this phenomenon, without providing a comprehensive understanding (*8, 9*). For instance, the critical current density (CCD) is commonly used to determine whether dendritic growth will occur (*10, 11*); in other words, it is considered that Li dendrites will only grow above the CCD. However, there is evidence that Li can repair itself and inhibit dendrite formation when the current density is sufficiently high (*12*). Additionally, it is widely accepted that the dendrite diameter increases with increasing temperature; however, self-healing sometimes occurs when the current density is greater than ~9 mA $cm^{-2}$ or when the temperature is ~60 °C (*12, 13*). These phenomena are difficult to explain and require further study.

The Li-metal deposition mechanism has been studied using several numerical simulation methods, including finite element simulations (*14-16*), ab initio calculations (*17-19*), and molecular dynamics simulations (*12, 20, 21*). Although finite element simulation can be performed at large scale and with long simulation time, it cannot reflect the movement of Li atoms during deposition due to its reliance on partial differential equations (*14*). Additionally, while ab initio calculations accurately reflect the atomic-scale kinetic and thermodynamic properties of Li (*18*), their high computational cost limits the simulation scale to hundreds of atoms and several picoseconds (*22*). Consequently, they are not suitable for analysing the emergence phenomenon of Li deposition, such as dendrite formation. Molecular dynamics simulations overcome the limitations of ab initio calculations due to the larger scale and higher speed; however, discrepancies remain between predictions based on classical Li potentials and experimental results (*23, 24*). Meanwhile, machine learning is an emerging and powerful technology that can be used to optimise potential function parameters with an accuracy approaching that of quantum-mechanics computations (*25-27*). Recently, machine-learning based potentials were successfully applied to several systems, including $N_2O_5$ in atmospheric aerosol (*28*), methane combustion (*29*), and disordered silicon (*30*).

Herein, we develop a Li potential model based on a deep potential neural network to achieve a large-scale (>100,000 atoms) simulation. The model predictions are close to those of ab initio calculations and experimental measurements. Two types of self-healing mechanisms are revealed: surface and bulk self-healing. We further identify three different Li-dendrite shapes: 'needle', 'mushroom', and 'hemisphere'. We finally introduce the concept of local current density (LCD) as a supplement to CCD, which compensates for experimentally observed contradictions. Homogeneous LCD distribution and high Li fluidity are essential for triggering the two self-healing processes.

**Surface self-healing**

We developed a Li surface deep potential model using a machine learning algorithm with the same accuracy as ab initio calculations during robust testing (see Methods, Supplementary Figs. 1–9, and Supplementary Table 1 for details). The melting point of Li predicted by the Li surface deep potential model is 451.6 K, which is close to the experimental value of 454 K. Furthermore, since the surface morphology is critical for Li deposition (*15, 18, 31*), we studied the dynamic behaviour of Li deposition with different initial atomic-scale surface morphologies (see Supplementary Information and Supplementary Fig. 10). Four surface morphologies were constructed for the initial molecular dynamics structures: flat surface, rectangular surface, positive triangular surface, and inverted triangular surface, as shown in the top panels of Fig. 1a–d, respectively. Interestingly, all the surface defects were gradually filled during the homogeneous deposition process to form smooth surfaces similar to that of the flat surface (Fig. 1b–d and Supplementary Video 1). Notably, despite the different surface defects, each surface first evolved to have a corrugated shape after ~0.3 ns, which is indicative of a more stable surface state. To verify this, we relaxed the rectangular configuration at 300 K without depositing Li atoms and noted that the relaxed surface structure also had a corrugated morphology (Supplementary Fig. 11). Further research on the effects of temperature ($T$), generation rate ($R_g$), and supercell shape on homogeneous deposition revealed that they all negligibly influenced the final morphology (Supplementary Information and Supplementary Figs. 12–15).

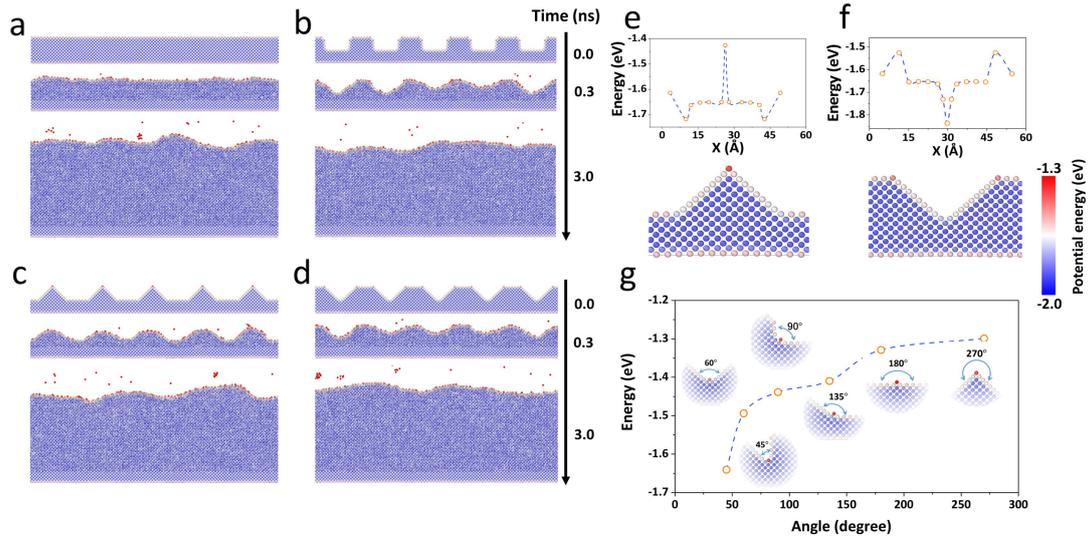

**Fig. 1. Li homogeneous deposition and surface self-healing.** Snapshots at 0, 0.3, and 3.0 ns when the initial surface had (**a**) a smooth configuration, (**b**) rectangular defects, (**c**) triangular defects, and (**d**) inverted triangular defects. Potential energy distribution of Li atoms with (**e**) upright triangular configuration, (**f**) inverted triangular configuration, and (**g**) valley configuration with six different angles (45°–270°). Different coloured balls correspond to different potential energies, as indicated by the colour scale bar.

The results presented above suggest that Li metal has inherent surface self-healing properties during homogeneous deposition. Hence, to clarify the surface self-healing mechanism, we calculated the atomic potential energies of the different constructions. We noted that the potential energy of the Li atoms in the bulk was generally lower than that on the surface, which indicates that the surface atoms are more active (Fig. 1e and f). When only the surface atoms are compared, the potential energy is higher for the atoms at the tips (−1.53 and −1.43 eV) and lower in the valleys (−1.84 and −1.72 eV in the upper subgraph of Fig. 1e and f, respectively) of the triangular surface. The primary difference between the tips and valleys is the angle formed by the local atoms. Therefore, to explore the effect of the angle on the potential energy, we placed a Li atom in a valley with various angles, which revealed that the potential energy increases with increasing angle (Fig. 1g). When a Li atom is placed on a sloped surface near the bottom of the valley, it relaxes to the bottom and remains stable (Supplementary Video 2). These results indicate that the surface self-healing mechanism originates in the varying potential energies of the Li atoms at various surface morphologies. This ability to self-heal is inherent to Li, which facilitates its growth into a perfect metal with a smooth dendrite-free surface under homogeneous deposition conditions.

**Bulk self-healing**

Inhomogeneous deposition is electrochemically favoured at high current densities, which leads to Li-dendrite growth (*3*). This phenomenon is usually considered to be related to the tip effect in the electric field or an uneven electrode surface (*15*). Hence, we simulated inhomogeneous deposition using a Li surface deep potential model to study the mechanism of dendrite growth and thereby determine a means to suppress it (see Methods and Supplementary Fig. 16). The results revealed that two hemispherical dendrites grow separately (300 ps in Fig. 2a), until they encounter each other (at ~660 ps in Fig. 2a) to form a larger hemispherical dendrite (at ~800 ps in Fig. 2a). Additionally, a cavity forms between the two dendrites after ~660 ps, which shrinks and then completely disappears as the two dendrites fuse at ~800 ps. The large hemisphere fuses with other symmetrical hemispheres with further deposition to finally form a smooth surface (at ~2,540 ps in Fig. 2a). This phenomenon is different to surface self-healing and can be referred to as 'bulk self-healing'.

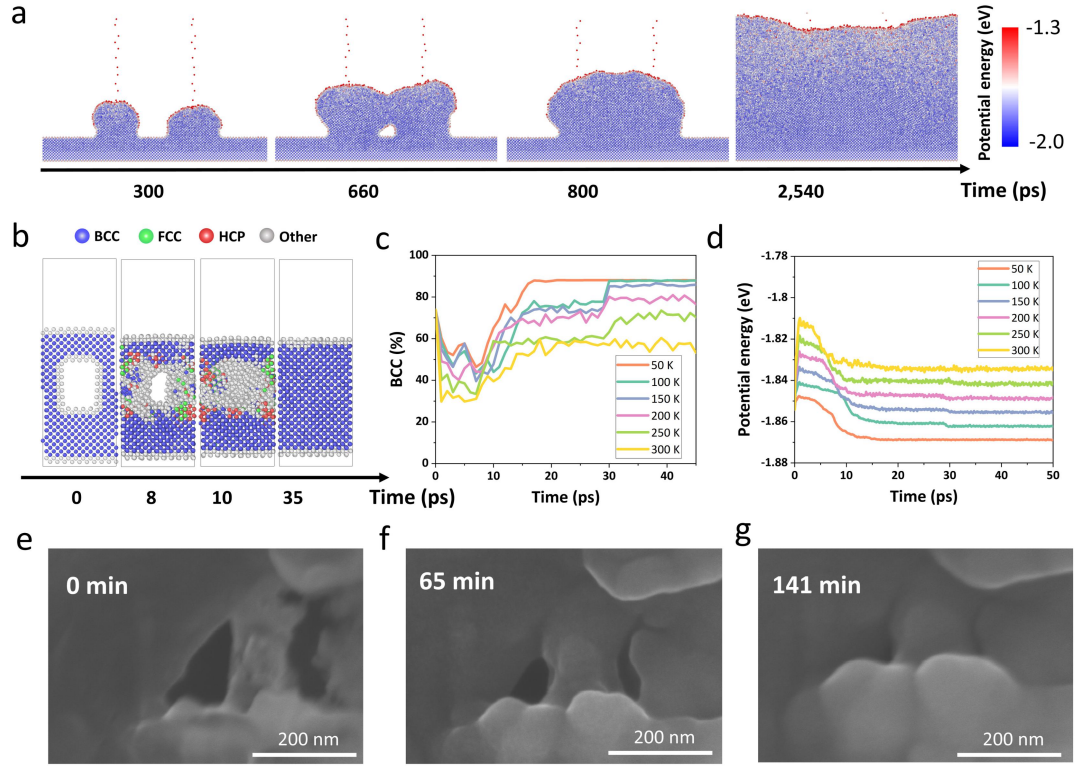

**Fig. 2. Inhomogeneous deposition and bulk self-healing.** (**a**) Snapshots at 300–2,540 ps during inhomogeneous deposition. (**b**) Bulk self-healing process at 50 K; changes in (**c**) body-centred cubic (BCC) proportion by adaptive common neighbour analysis and (**d**) potential energy with deposition time at different temperatures during bulk self-healing. SEM images of defects in the Li metal surface (**e**) 0 min, 65 min (**f**) and 141 min (**g**) after the defects were created. The ball colours in (**a**) denote the potential energy and correspond to the colour scale bar. The blue, green, and red balls in (**b**) denote atoms with local BCC, face-centred cubic (FCC), and hexagonal close-packed (HCP) arrangements, respectively, and the grey balls denote other local environments, including surface and amorphous atoms.

To further study the bulk self-healing mechanism, we simulated the growth of a Li crystal with a cuboid cavity in the temperature range of 50–300 K (Fig. 2b), which revealed that the cavity gradually shrinks until it completely disappears at ~30 ps, regardless of temperature; adaptive common neighbour analysis (a-CNA) (*32*) revealed that body-centred cubic (BCC) states first decrease in number and then increase until self-healing is complete (Fig. 2c). This suggests that some of the Li phase transforms to the amorphous state during deposition and then back to the crystalline state after self-healing. Amorphous Li, which is more fluid than crystalline Li, as characterised by mean squares displacement (Supplementary Fig. 17), is formed during the self-healing process at 8–10 ps (Fig. 2b), which results in cavity shrinkage. The change in the potential energy of the Li atoms and the bulk self-healing process are depicted in detail in Supplementary Video 3. The average potential energy of the Li atoms decreases during the process via an apparently barrierless mechanism, which is the main driving force for cavity shrinkage (Fig. 2d); the initial energy increases at ~1 ps owing to the increase in temperature from 0 K to the target value. Amorphous Li has a lower potential energy than that of surface Li and is more fluid than crystalline Li even at a low temperature (Supplementary Fig. 17). Amorphous Li acts as a lubricative intermediate that triggers bulk self-healing, resulting in perfect Li metal. However, it is worth mentioning that this bulk self-healing phenomenon was not observed in a recent Li-deposition simulation study (*21*). The discrepancy is possibly attributable to the different Li potential used (a classical potential for Li metal) or the quasi-three-dimensional configuration.

In order to experimentally validate the simulation results, we conducted Li bulk self-healing experiments (see Methods). Scanning electron microscope (SEM) images show that after we created defects with diameters of ~100 nm (Fig. 2e), the diameter of the defects continuously decreased, and their edges became smoother (Fig. 2f). The defects disappeared completely after 141 min (Fig. 2g). We recorded the SEM images at room temperature (~26 °C) and strictly controlled the power and exposure time of the electron beam on the defect of the sample to ensure that the Li foil self-healed at a temperature far below the melting temperature. This self-healing was also observed in other Li metal defects (Supplementary Fig. 18), indicating the universality of this phenomenon. These experiments thus demonstrated that Li metal can indeed experience bulk self-healing at the nanoscale, thereby validating the predictions of our simulation.

**Dendrite morphology**

Inhomogeneous deposition with the tip effect was simulated at three different temperatures ($T$ = 200, 300, and 400 K), and generation rates ($R_g$ = 0.5, 1, and 5 Li ps$^{-1}$) (see Supplementary Information and Methods for simulation conditions). Figure 3 shows snapshots of the dendrite at an altitude of 10 nm at various temperatures and generation rates (in contrast, snapshots of the dendrite with ~6,000 deposited Li atoms are shown in Supplementary Fig. 19). The bottom diameter of the dendrite is seen to increase with temperature at a constant generation rate, while the top diameter increases with increasing generation rate at a constant temperature. At a high generation rate (5 Li ps$^{-1}$) (Fig. 3g–i), a hemispherical dendrite is formed, the bottom of which has noticeable lattice fringes indicative of crystal formation, while the top part remains amorphous. Calculations of the local dendrite temperature revealed that the temperatures at the bottom and top of the dendrite are significantly different (Supplementary Fig. 20); the temperature at the bottom was ~300 K, while that at the surface may have been ~800 K when the temperature was maintained at 300 K (Supplementary Fig. 20). In contrast, at a low generation rate, the temperature at the top is much closer to that at the bottom. For example, at 0.5 Li ps$^{-1}$, the bottom and top temperatures are ~300 and 363 K, respectively (Supplementary Fig. 20). Generally, the high temperature at the dendrite surface is attributable to the energy released by the condensation of the deposited Li atoms, especially at a high generation rate (see Supplementary Information for further discussion).

Based on the simulation results presented above, we conclude that temperature is an important factor in the deposition process because it controls the fluidities of the bottom and top (surface) atoms, which determine the dendrite morphology. To illustrate this point, the mean squares displacement values of Li on the surface and in the bulk were calculated with temperatures in the range of 50–450 K (Supplementary Fig. 21), showing that the surface Li atoms are more fluid than those in the bulk, regardless of temperature. All Li atoms are nearly immobile at lower temperatures, whereas the surface atoms are slightly more fluid. Furthermore, the diffusion coefficient of Li on the surface is nearly three orders of magnitude higher than that of the bulk (Supplementary Fig. 22). Therefore, the above results confirm that the difference in the fluidities of the bottom and top (surface) atoms is critical to the morphology of the dendrite.

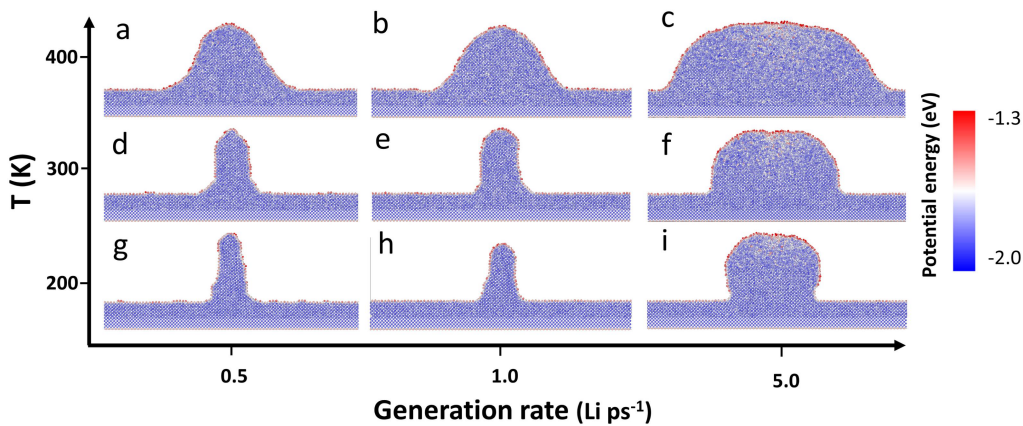

**Fig. 3.** Snapshots of inhomogeneous deposition at different temperatures (*T*) and generation rates ($R_g$) at a dendrite altitude of 10 nm. $R_g$ = 0.5 Li ps$^{-1}$ (**a, d, g**), 1.0 Li ps$^{-1}$ (**b, e, h**), and 5.0 Li ps$^{-1}$ (**c, f, i**). *T* = 400 K (**a–c**), 300 K (**d–f**), and 200 K (**g–i**).

Dendrite morphologies can be divided into three types: 'needle', 'mushroom', and 'hemisphere'. At a low temperature (e.g. 200 K) and low generation rate, the surface Li atoms are poorly mobile, leading to a 'needle' morphology in the deposited area (Fig. 3d and g). The deposited Li atoms immediately crystallise upon contact with the Li-metal surface, resulting in a clear lattice fringe (Fig. 3g and h). During deposition, the dendrite frequently fractures because Li atoms constantly collide with the surface, resulting in kinks in the dendrite (Supplementary Fig. 19g and Supplementary Video 4), which is consistent with previous observations (*33, 34*). This dendrite fracturing behaviour may explain the branched Li-dendrite morphology observed in these experiments (*14*), considering that the tips are expected to develop into other dendrites due to the tip effect. The fluidity of the surface Li atoms increases with increasing temperature, leading to a larger area for Li-atom diffusion; hence, the 'needles' have larger diameters at higher temperatures (Fig. 3a–c).

At a low temperature and high generation rate, the dendrite favours a 'mushroom' morphology owing to the significant difference in the fluidities of the bottom and top atoms (Fig. 3i and Supplementary Fig. 19i). This morphology is attributable to the fact that crystallisation occurs easily at the bottom of the dendrite, leading to poor Li fluidity, while the higher surface temperature at the tip results in rapid Li diffusion that forms a 'mushroom' morphology.

At a high temperature and high generation rate, Li exists on the surface in a super-fluid liquid state, leading to a 'hemispherical' dendrite (Fig. 3c). Contact between two 'hemispherical' dendrites triggers the bulk self-healing mechanism, which suppresses dendrite formation (Fig. 2a and Supplementary Video 5). Therefore, increasing the fluidity of the surface Li atoms by increasing the temperature and generation rate is an effective method for suppressing dendrite growth.

Our conclusions may be more applicable to solid electrolyte conditions without solid electrolyte interphase (SEI) formation, which may prevent dendrite contact; however, dendrites can fuse at high temperature with some special SEIs (*13*). The mechanism responsible for fusion still warrants further study with respect to the composition, thickness, elasticity, and other SEI factors. While some solid electrolytes have higher shear moduli than Li, studies suggest that they are still unable to inhibit dendrite growth, especially in materials with high ionic conductivities (*35*). The surface liquid state of Li may significantly contribute to this problem, because a high shear modulus can prevent solid Li-dendrite formation (*2*), but cannot prevent liquid Li from flowing in the grain boundaries.

**Discussion**

Although many studies have suggested that the CCD is a crucial factor for dendrite growth (*11, 35, 36*), we hypothesise that it is not the decisive factor because it is an average of the current density; for example, a subregion may have a high current density (e.g. tip effect or poor contact) even at a low CCD. Therefore, we further suggest that the LCD distribution and variance in LCD (VLCD) are key factors that affect dendrite growth. For example, Li is mainly deposited homogeneously when the VLCD is close to zero, which triggers surface and bulk self-healing (Fig. 4a). Self-healing continues to smoothen the surface at high CCD and low VLCD (Fig. 4b), especially in the solid electrolyte system without the SEI (*37*). In contrast, inhomogeneous deposition will dominate at a relatively large VLCD, which prevents surface self-healing (Fig. 4c). In this case, the dendrite morphology and the distance between the dendrites are the main factors that affect bulk self-healing. The larger dendritic 'hemispheres' can easily encounter each other, which triggers bulk self-healing and results in a smooth surface (Supplementary Fig. 23 and Supplementary Video 5). We expect that the same process occurs in the case of the 'mushroom' dendrites (Supplementary Fig. 23 and Supplementary Video 6); however, it is difficult for the 'needle' dendrites to encounter each other, which hinders self-healing (Supplementary Fig. 23).

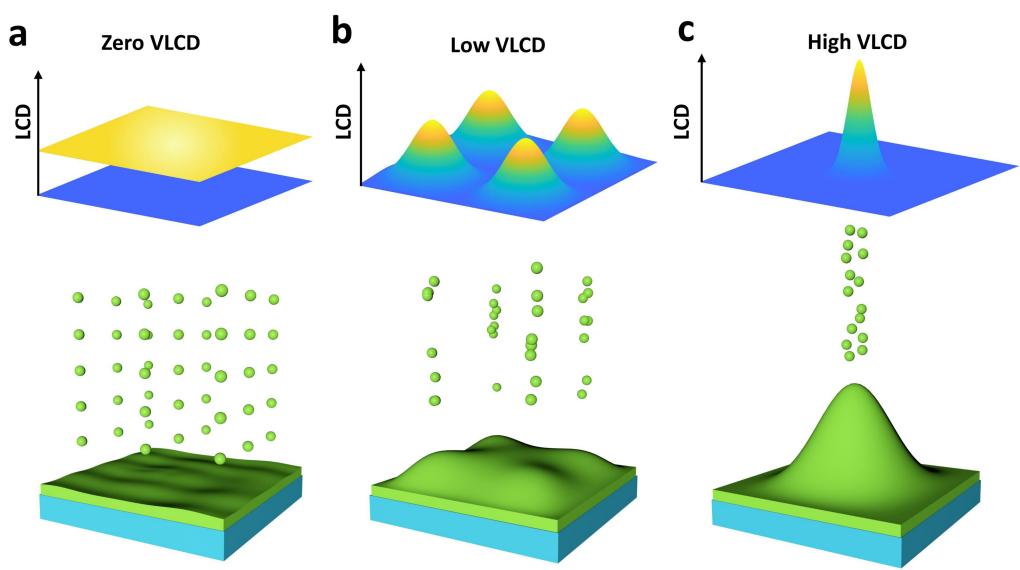

**Fig. 4. Schematic of Li deposition at different variance of local current density**

**(VLCD). (a)** Homogeneous deposition with VLCD close to zero, (**b**) inhomogeneous deposition with low VLCD, and (**c**) inhomogeneous deposition with high VLCD.

Many previously reported results were consistent with our simulations. An investigation (*12*) into the cycling performance of Li batteries at different current densities revealed that Li develops a smoother surface at a high current density (above ~9 mA cm$^{-2}$). The high current density results in a high dendrite temperature, which triggers extensive surface and bulk self-healing, thereby smoothening the dendrites and enhancing the cycling performance of the battery. Researchers have also found that the Li nuclei increase in size with increasing deposition temperature or current density, which is in good agreement with our simulation results (*12, 13*). In addition to the influence of temperature, the time required for Li self-healing is also related to the size of the defect. In our simulation, the defect size (diameter) of the Li is only ~1 nm, resulting in the self-healing process being finished in a few nanoseconds. In our experiments, the defect size of Li metal is ~100 nm, resulting in hundreds of minutes to finish the healing. In the previous report, pores between the dendrites were usually several microns, which led to the fusion of the dendrites took up to 3 days, even at 70 °C (*12*). Recently, a long-cycle-life all-solid-state Li-metal battery enabled through the use of an Ag-C composite was reported (*37*). The researchers attributed the excellent performance to the uniform deposition. We infer that the ultra-fast self-healing of tiny defects made by the uniform deposition is also key to forming the dendrite-free deposition morphology.

Based on the above discussion, several strategies can be adopted to trigger self-healing and improve the performance of Li-metal batteries. First, the fluidity of the deposited Li atoms can be improved, which is easily achieved by increasing the temperature or introducing amorphous lithium. Since crystalline (BCC) Li always exhibits an inferior fluidity, we speculate that introducing defects can significantly improve the fluidity of the deposited Li. Besides, doping with elements like silver and caesium may also improve the Li self-healing capability (*31, 37*). Second, a homogeneous deposition environment should be created. Deposition areas can be enlarged using three-dimensional clutter (*38*) or by smoothing the Li surface, thereby increasing the LCD and lowering the VLCD. Third, a solid electrolyte or artificial SEI can be used (*39*) to avoid SEI formation between the Li dendrites, which effectively facilitates dendritic bulk self-healing. Furthermore, a comprehensive plan involving all the above strategies may be more efficient because they are not mutually exclusive.

**Conclusion**

In summary, high fluidity (e.g. increasing $T$ or $R_g$) and low VLCD are key to inhibiting dendrite growth. High fluidity enlarges the dendrite, which facilitates contact and triggers surface and bulk self-healing. A low VLCD is indicative of a more uniform distribution in the anode, which effectively triggers surface and bulk self-healing. Thus, charging the Li battery at a relatively high temperature (350 K) and modifying the current collector to one with a more homogeneous surface, especially in a solid electrolyte system, can potentially suppress dendrite growth. This study introduces a promising strategy for investigating the Li-deposition mechanism at the atomic level using a machine learning potential, and provides new perspectives with respect to future research into improving the performance of Li-metal anodes.

**Acknowledgements**

We thank Gaoxiang Sun, Kuang Yu and Yancong Feng for technical support. This work was supported by the Starting Fund of Peking University, Shenzhen

## Methods

**DFT calculations.** We used the Vienna ab initio simulation package (VASP) (*40, 41*) based on density functional theory (DFT) to generate the training and testing datasets. The generalized gradient approximation (GGA) with a parametrized exchange-correlation function according to Perdew Burke and Ernzerhof (PBE) was used during calculations (*42*). The valence electron wave functions were expanded in the plane wave basis sets, and the projector augmented wave (PAW) method was used to describe the core-electron interactions (*43*). The plane-wave cut-off energy was set to be 380 eV for all calculations after a cut-off energy test range from 50 eV to 800 eV. This cut-off value is sufficient to reproduce accurate results for lithium and has a relatively low cost of computing resources. About 10,000 bulk configurations and 6,000 surface configurations were selected for the model training (see Supplementary Information for DFT calculations and datasets details).

**Training models.** We used the neural network proposed by Zhang et al. (*26, 44-46*) to develop the Li potential models. In this method, the total energy of the system is the sum of all the single atomic energies: $E = \Sigma E_i$. The energy of each atom ($E_i$) is determined by its local environment (other atoms) within a cut-off radius, defined as $E_i = \varphi_\omega(D_i)$; where $\varphi$ is the function of the neural network, $\omega$ is the parameter of the networks, and $D_i$ is the descriptor of the local environment for the $i$th atom. The diversity of configuration space is the key to an effective training model. In our work, we adopted an active learning scheme (*45-47*) to generate various configurations and regenerate an effective training data set (see Supplementary Information for more detail).

**Molecular Dynamics Simulations.** All the molecular dynamics simulations were conducted using LAMMPS (*48*). All the configurations used in Li deposition were supercells of the primitive BCC cell with an experimental lattice constant of 3.51 Å (*23, 49*). In the deposition simulation, eighteen layers of Li atoms are the substrate with the Miller indices of (100), and eight bottom layers of atoms were fixed for a bulk environment. The NVT ensemble with the Langevin thermostat was used in all deposition simulations. A time step of 1 fs was used in all the simulations. All visualizations of the molecular dynamics trajectory were performed with the OVITO program (*50*), and the adaptive common neighbour analysis (a-CNA) method (*32*) was used to identify and distinguish a typical phase.

**Homogeneous deposition simulation conditions.** The lattice constant of the supercells (Supplementary Fig. 10) used in this simulation is 28.07 nm × 1.40 nm × 30.00 nm. Li atoms are generated at the top of the supercell, and the probability of atom generation at X and Y coordinates obeys the distribution of X~U (0, 28.07) and Y~U (0, 1.40), where U stands for uniform distribution. The time interval of atom generation (generation rate) is 5 Li ps$^{-1}$, and the falling speed of lithium atoms is 500 m s$^{-1}$. These simulations were performed at 300 K by NVT ensemble with a total deposition time of 3 ns.

**Inhomogeneous deposition simulation conditions.** The supercells in inhomogeneous deposition have the same dimensions as those in homogeneous deposition (Supplementary Fig. 16, Fig. 2a, and Fig. 3). Lithium atoms were generated at the top of the supercell. The generation region was limited to two small regions in Supplementary Fig. 16 and Fig. 2a, and one small region in Fig. 3. The generation rate and the falling rate were the same as the homogeneous deposition conditions shown in Fig. 1a. The simulation, whose results are depicted in Fig. 2a, was performed at 100 K by NVT ensemble with a total deposition time of 3 ns. In the simulation, whose results are shown in Fig. 3, the falling speed of Li is proportional to the generation rate. For example, when the generation rate is 0.5 Li ps$^{-1}$, the falling speed is 500 m s$^{-1}$, and when the generation rate is 1 Li ps$^{-1}$, the falling speed is 1000 m s$^{-1}$, to maintain a constant distance between the two falling atoms.

**Experiments.** The Li metal foil with a fresh and smooth surface (without the passivation layer) was prepared in the glove box with water content less than 0.1 ppm and oxygen content less than 0.1 ppm. The defects on the Li foil were created with a tiny needle. The diameter of the defects was ~100 nm. Then the Li foil was transferred to the scanning electron microscope (SEM) sample box without any exposure to the air. The defects made by the needle were searched and characterized under the SEM. The exposure time of Li foil for each image was about 5-8 seconds. The electron beam was quickly moved away from the characterized areas after each image was taken to reduce the influence of electron beam irradiation on the Li surface.

The whole implementation process was carried out at room temperature, and the Li foil did not melt during the measurement. All images were taken at regular intervals to observe the self-healing process of the Li defects.

## Data and materials availability

All the data are available in the main text or the supplementary materials.

## References and Notes

1. J. M. Tarascon, M. Armand, Issues and challenges facing rechargeable lithium batteries. *Nature* **414**, 359-367 (2001).

2. W. Xu *et al.*, Lithium metal anodes for rechargeable batteries. *Energy & Environmental Science* **7**, 513-537 (2014).

3. B. Liu, J.-G. Zhang, W. Xu, Advancing Lithium Metal Batteries. *Joule* **2**, 833-845 (2018).

4. X.-B. Cheng, R. Zhang, C.-Z. Zhao, Q. Zhang, Toward Safe Lithium Metal Anode in Rechargeable Batteries: A Review. *Chemical Reviews* **117**, 10403-10473 (2017).

5. P. P. Paul *et al.*, A review of Existing and Emerging Methods for Lithium Detection and Characterization in Li-Ion and Li-Metal Batteries. *Adv. Energy Mater.* **11**, 2100372 (2021).

6. Y. He *et al.*, Origin of lithium whisker formation and growth under stress. *Nature nanotechnology* **14**, 1042-1047 (2019).

7. S. Wang, H. Xu, W. Li, A. Dolocan, A. Manthiram, Interfacial Chemistry in Solid-State Batteries: Formation of Interphase and Its Consequences. *Journal of the American Chemical Society* **140**, 250-257 (2018).

8. Y. Sun *et al.*, Boosting the Optimization of Lithium Metal Batteries by Molecular Dynamics Simulations: A Perspective. *Advanced Energy Materials* **10**, 2002373 (2020).

9. P. Albertus *et al.*, Challenges for and Pathways toward Li-Metal-Based All-Solid-State Batteries. *ACS Energy Lett.* **6**, 1399-1404 (2021).

10. F. Han *et al.*, High electronic conductivity as the origin of lithium dendrite formation within solid electrolytes. *Nature Energy* **4**, 187-196 (2019).

11. Y. Lu *et al.*, Critical Current Density in Solid-State Lithium Metal Batteries: Mechanism, Influences, and Strategies. *Adv. Funct. Mater.* **31**, 2009925 (2021).

12. L. Li *et al.*, Self-heating–induced healing of lithium dendrites. *Science* **359**, 1513-1516 (2018).

13. J. Wang *et al.*, Improving cyclability of Li metal batteries at elevated temperatures and its origin revealed by cryo-electron microscopy. *Nature Energy* **4**, 664-670 (2019).

14. A. Jana, S. I. Woo, K. S. N. Vikrant, R. E. García, Electrochemomechanics of lithium dendrite growth. *Energy & Environmental Science* **12**, 3595-3607 (2019).

15. S.-H. Wang *et al.*, Stable Li Metal Anodes via Regulating Lithium Plating/Stripping in Vertically Aligned Microchannels. *Advanced Materials* **29**, 1703729 (2017).

16. J. Wu *et al.*, Polycationic Polymer Layer for Air-Stable and Dendrite-Free Li Metal Anodes in Carbonate Electrolytes. *Adv. Mater.* **33**, 2007428 (2021).

17. M. Jäckle, A. Groß, Microscopic properties of lithium, sodium, and magnesium battery anode materials related to possible dendrite growth. *The Journal of Chemical Physics* **141**, 174710 (2014).

18. M. Jäckle, K. Helmbrecht, M. Smits, D. Stottmeister, A. Groß, Self-diffusion barriers: possible descriptors for dendrite growth in batteries? *Energy Environ. Sci.* **11**, 3400-3407 (2018).

19. J. Jiao, R. Xiao, M. Han, Z. Wang, L. Chen, Impact of hydrogen on lithium storage on graphene edges. *Applied Surface Science* **515**, 145886 (2020).

20. M. Yang, Y. Liu, A. M. Nolan, Y. Mo, Interfacial Atomistic Mechanisms of Lithium Metal Stripping and Plating in Solid-State Batteries. *Advanced Materials* **33**, 2008081 (2021).

21. X. Wang *et al.*, Glassy Li metal anode for high-performance rechargeable Li batteries. *Nature Materials* **19**, 1339-1345 (2020).

22. J. Behler, First Principles Neural Network Potentials for Reactive Simulations of Large Molecular and Condensed Systems. *Angew. Chem. Int. Ed.* **56**, 12828-12840 (2017).

23. A. Nichol, G. J. Ackland, Property trends in simple metals: An empirical potential approach. *Phys. Rev. B* **93**, 184101 (2016).

24. J. R. Vella, F. H. Stillinger, A. Z. Panagiotopoulos, P. G. Debenedetti, A Comparison of the Predictive Capabilities of the Embedded-Atom Method and Modified Embedded-Atom Method Potentials for Lithium. *The Journal of Physical Chemistry B* **119**, 8960-8968 (2015).

25. J. Behler, M. Parrinello, Generalized Neural-Network Representation of

High-Dimensional Potential-Energy Surfaces. *Physical Review Letters* **98**, 146401 (2007).

26. L. Zhang, J. Han, H. Wang, R. Car, W. E, Deep Potential Molecular Dynamics: A Scalable Model with the Accuracy of Quantum Mechanics. *Physical Review Letters* **120**, 143001 (2018).

27. A. P. Bartók, M. C. Payne, R. Kondor, G. Csányi, Gaussian Approximation Potentials: The Accuracy of Quantum Mechanics, without the Electrons. *Physical Review Letters* **104**, 136403 (2010).

28. M. Galib, D. T. Limmer, Reactive uptake of N2O5 by atmospheric aerosol is dominated by interfacial processes. *Science* **371**, 921-925 (2021).

29. J. Zeng, L. Cao, M. Xu, T. Zhu, J. Z. H. Zhang, Complex reaction processes in combustion unraveled by neural network-based molecular dynamics simulation. *Nature Communications* **11**, 5713 (2020).

30. V. L. Deringer *et al.*, Origins of structural and electronic transitions in disordered silicon. *Nature* **589**, 59-64 (2021).

31. F. Ding *et al.*, Dendrite-Free Lithium Deposition via Self-Healing Electrostatic Shield Mechanism. *J. Am. Chem. Soc.* **135**, 4450-4456 (2013).

32. A. Stukowski, Structure identification methods for atomistic simulations of crystalline materials. *Modelling and Simulation in Materials Science and Engineering* **20**, 045021 (2012).

33. H. Ghassemi, M. Au, N. Chen, P. A. Heiden, R. S. Yassar, Real-time observation of lithium fibers growth inside a nanoscale lithium-ion battery. *Applied Physics Letters* **99**, 123113 (2011).

34. A. Kushima *et al.*, Liquid cell transmission electron microscopy observation of lithium metal growth and dissolution: Root growth, dead lithium and lithium flotsams. *Nano Energy* **32**, 271-279 (2017).

35. Z. Lu *et al.*, Modulating Nanoinhomogeneity at Electrode–Solid Electrolyte Interfaces for Dendrite-Proof Solid-State Batteries and Long-Life Memristors. *Adv. Energy Mater.* **11**, 2003811 (2020).

36. D. Lin, Y. Liu, Y. Cui, Reviving the lithium metal anode for high-energy batteries. *Nature Nanotechnology* **12**, 194-206 (2017).

37. Y.-G. Lee *et al.*, High-energy long-cycling all-solid-state lithium metal batteries enabled by silver–carbon composite anodes. *Nat. Energy* **5**, 299-308 (2020).

38. C.-P. Yang, Y.-X. Yin, S.-F. Zhang, N.-W. Li, Y.-G. Guo, Accommodating lithium into 3D current collectors with a submicron skeleton towards long-life lithium metal anodes. *Nat. Commun.* **6**, 8058 (2015).

39. Y. Liu *et al.*, An Artificial Solid Electrolyte Interphase with High Li-Ion

Conductivity, Mechanical Strength, and Flexibility for Stable Lithium Metal Anodes. *Adv. Mater.* **29**, 1605531 (2017).

40. G. Kresse, J. Furthmüller, Efficient iterative schemes for ab initio total-energy calculations using a plane-wave basis set. *Physical Review B* **54**, 11169-11186 (1996).

41. G. Kresse, J. Furthmüller, Efficiency of ab-initio total energy calculations for metals and semiconductors using a plane-wave basis set. *Computational Materials Science* **6**, 15-50 (1996).

42. J. P. Perdew, K. Burke, M. Ernzerhof, Generalized Gradient Approximation Made Simple. *Physical Review Letters* **77**, 3865-3868 (1996).

43. P. E. Blöchl, Projector augmented-wave method. *Physical Review B* **50**, 17953-17979 (1994).

44. H. Wang, L. Zhang, J. Han, W. E, DeePMD-kit: A deep learning package for many-body potential energy representation and molecular dynamics. *Computer Physics Communications* **228**, 178-184 (2018).

45. L. Zhang, D.-Y. Lin, H. Wang, R. Car, W. E, Active learning of uniformly accurate interatomic potentials for materials simulation. *Physical Review Materials* **3**, 023804 (2019).

46. Y. Zhang *et al.*, DP-GEN: A concurrent learning platform for the generation of reliable deep learning based potential energy models. *Computer Physics Communications* **253**, 107206 (2020).

47. J. Wu, Y. Zhang, L. Zhang, S. Liu, Deep learning of accurate force field of ferroelectric HfO2. *Physical Review B* **103**, 024108 (2021).

48. S. Plimpton, Fast Parallel Algorithms for Short-Range Molecular Dynamics. *Journal of Computational Physics* **117**, 1-19 (1995).

49. W.-S. Ko, J. B. Jeon, Interatomic potential that describes martensitic phase transformations in pure lithium. *Computational Materials Science* **129**, 202-210 (2017).

50. A. Stukowski, Visualization and analysis of atomistic simulation data with OVITO-the Open Visualization Tool. *Modelling and Simulation in Materials Science and Engineering* **18**, 015012 (2010).

**Supplementary Information**

Materials and Methods

Supplementary Figs. 1 to 23

Supplementary Table 1

References

Supplementary Videos 1 to 6

# Supplementary Information for

# Self-healing mechanism of lithium in lithium metal batteries

Junyu Jiao[1*], Genming Lai[1*], Liang Zhao[2,5], Jiaze Lu[3], Qidong Li[2,5], Xianqi Xu[1], Yao Jiang[4], Yan-Bing He[2], Chuying Ouyang[4], Feng Pan[1], Hong Li[3†], and Jiaxin Zheng[1†]

[*]These authors contributed equally to this work.
Correspondence to: zhengjx@pkusz.edu.cn; hli@iphy.ac.cn;

**Contents：**
1. Supplementary Methods
2. Supplementary Figures (Figs. S1 to S22)
3. Supplementary Table S1
4. Supplementary References

# Supplementary Methods

## S1. Model Generation and Test

**DFT calculations and datasets.** We used the Vienna ab initio simulation package (VASP) [1, 2] based on density functional theory (DFT) to generate the training and testing datasets. The generalized gradient approximation (GGA) with a parametrized exchange-correlation function according to Perdew Burke and Ernzerhof (PBE) was used during the calculations [3]. The valence electron wave functions were expanded in the plane wave basis sets, and the projector augmented wave (PAW) method was used to describe the core-electron interactions [4]. The plane-wave cutoff energy was set to be 380 eV for all calculations after a cutoff energy test range from 50 eV to 800 eV. This cutoff value is sufficient to reproduce the accurate results for lithium and has a relatively low cost of computing resources.

To have a reliable prediction of the lithium surface properties, we generated two data sets: bulk data and surface data. The basic structure we used in the bulk data is a supercell of 4×4×4 larger than the primitive bcc cell. The primary surface structures contained three low Miller index surfaces, which are (1 0 0), (1 1 0), and (1 1 1), respectively. The surface structures were terminated by more than a 20 Å vacuum interval. The Brillouin zone samplings were performed using a 3×3×3 k-point grid in the Monkhorst-Pack scheme for the bulk structure and 3×3×1 for the surface structures. The convergence precision of electron self-consistent field calculation was less than $10^{-7}$ eV. Totally, about 10,000 bulk configurations and 6,000 surface configurations were selected for the model training.

The ab initio molecular dynamics (AIMD) generated the testing data sets in a canonical ensemble (NVT) with temperatures ranging from 200 K to 1,000K. Three supercells of 4×4×4 larger than the primitive bcc cell with randomly disturbed were used as the initial configurations. The total time steps of AIMD were 15,000, and 1,500 configurations were selected for the bulk test dataset. The initial surface configurations contained three low Miller index surfaces, same with training datasets, with total AIMD time steps of 12,000 and 1,000 configurations were selected for the surface test dataset.

**Training models.** We used the neural network to develop the Li potential models proposed by Zhang et al [5, 6, 7, 8]. In this method, the total energy of the system is the sum of all the single atomic energy: $E = \Sigma E_i$. The energy of each atom ($E_i$) is determined by its local environment (other atoms) within a cutoff radius, defined as $E_i = \varphi_\omega(D_i)$, where $\varphi$ is the function of the neural network, $\omega$ is the parameter of the networks, and $D_i$ is the descriptor of the local environment for the ith atom. This local environment is usually complex, nonlinear, and has the characteristics of many-body interaction. However, the high-dimensional neural network has a robust fitting ability and can establish the map between the local atomic information and atomic energy (forces) through appropriate training.

The smooth version of the deep neural network model (DP-SE) was employed in our work through the DEEPMD-Kit Python package [6, 9]. The model includes two networks: the embedding network and the fitting network. Before the embedding network, a local coordinate is first established for each atom, and the position information of the local environment within the cutoff radius can be obtained. This position information includes the distance of the surrounding atoms to the target atom (reciprocal) in the three dimensions (reciprocal). This position information defined as $D_i$ are then inputted to the embedded network as the descriptor of the target atom for further processing to ensure the rotation, translation, displacement invariance, and continuity of DP-SE. The cutoff radius was set to 6.8 Å, and the descriptors decay smoothly from 5.8 Å to 6.8 Å. Both the embedding network and the fitting network use the ResNet-like architecture [10]. The loss function of DP-SE was defined as:

$$L = P_e \Delta E^2 + \frac{P_f}{3N}\sum_i |\Delta F_i|^2 + \frac{P_\xi}{9}||\Delta \xi||^2 \quad (1)$$

where Δ denotes the difference between the DP-SE prediction and the training data, $N$ is the number of atoms, $E$ is the energy per atom, $F_i$ is the atomic force of atom i, and ξ is the virial tensor divided by $N$. $P_e$, $P_f$, and $P_\xi$ are tunable prefactors. Here we increase both $P_e$ and $P_\xi$ from 0.02 to 1, and decreases $P_f$ from 1,000 to 2. The network parameters are updated by the backpropagation algorithm in training (the Adam optimization strategy is adopted [11]). The energy and force results with the decrease of L will be close to those calculated by the DFT.

**The Active Learning.** The diversity of configuration space is the key to an effective training model. In our work, we adopted an active learning scheme[7, 8, 12] to generate various configurations and regenerate an effective training data set. Active learning contains three steps, which are exploration, labeling, and training. These steps will repeat automatically until the accurate models are established (fig. S1). In the exploration step, we use four different deep neural network potential (DP) models (with varying initialization parameters) to calculate the energy and force for a configuration and compare the difference from these models. we define the model deviation ε as the maximal standard deviation of the atomic force predicted by the model ensemble:

$$\varepsilon = \max_i \sqrt{\langle ||f_i - \langle f_i \rangle||^2 \rangle} \quad (2)$$

Where $i$ runs through the atomic indices in a configuration, and the ensemble average ⟨..⟩ is taken over the ensemble of models. When a small ε value is reached, all the models give a uniform result for each interatomic force in the configuration. This indicates that the configuration is near the existing training set and can be accurately predicted by all the models. Thus, such configurations will not be labeled.

If ε is relatively large, it means that the model ensemble has not been effectively trained near the configuration (possibly due to the lack of diversity in the training data set). Therefore, it is necessary to label this configuration for DFT calculation and put it in the training data set. However, a much large ε may also mean that the configuration itself is unreasonable. Thus, we introduce two thresholds and only label the configuration when 0.02 eV/Å < ε < 0.2 eV/Å. For the labeled configurations, we performed DFT calculations to get accurate energy and force results. The new training dataset was added to the previous training datasets to retrain the models. In the next iteration of the training step, we adopted the transfer learning strategy. That is, the initial parameters are not initialized randomly, but the model parameters of the last training are used. The above three steps will be repeated until every ε of all configurations is less than 0.02 eV/ Å.

**Model test.** We developed a lithium deep surface potential model (Li-SP) by ML algorithm, realizing the accuracy of ab initio calculations. Meanwhile, Li-SP can achieve a large-scale simulation of over 100,000 atoms with linear computational costs as system size (using a GPU of NVIDIA Tesla V100 32G). To measure the accuracy of Li-SP, we first calculated the mean absolute error (MAE) of the energy and force between Li-SP and DFT calculation results, respectively, based on the test datasets (including bulk and surface datasets, see Supplementary

S2). The energy MAE is 0.85 meV/atom in the bulk test dataset and 1.21 meV/atom in the surface test dataset (fig. S2). The average MAEs of force in three dimensions are 0.019 eV Å$^{-1}$ in the bulk test dataset and 0.015 eV Å$^{-1}$ in the surface test dataset (fig. S3). The equation of state (EOS) of Li calculated by different methods (Li-SP, EAM[13], MEAM[14], and DFT) shows that the EOS of Li-SP are almost perfectly consistent with the EOS of DFT (fig. S4). By contrast, the EOS of EAM and MEAM show a significant deviation from the DFT results. Besides, the radial distribution function (RDF) of lithium at 700 K also shows that the RDF of Li-SP is consistent with the DFT results, more accurate than the results of MEAM and EAM (fig. S5). The melting point of lithium predicted by Li-SP in the solid-liquid mixed-phase (NVE ensemble) is 451.6 K (fig. S6), which is very close to the experimental value of 454 K [15]. From this point (table. S1), Li-SP is the most reliable potential among the Li potentials [13, 14, 16]. The mean square displacement (MSD) of Li at 1,000 K was also calculated by DFT and Li-SP, and no obvious difference between the two methods (fig. S7). Table S1 lists the predicted values of Li properties by different methods and the comparison with experimental results. It is remarkable that the error between Li-SP and DFT is the smallest in terms of surface energy. All the above results indicate that the Li-SP has the same accuracy as DFT calculations, and it can provide reliable results during the large-scale simulation of lithium deposition.

To highlight the accuracy of Li-SP in the calculation of surface configuration, we also tested Li bulk potential (Li-BP), which has similar networks with Li-SP but was trained without the surface training set. The results were shown in figs. S8 and S9 and Table S1. The MAE of surface energy test in Li-BP is 2.39 meV/atom (fig. S8), much higher than 1.21 meV/atom in Li-SP. The MAE of surface force test of Li-BP in the Z direction is 0.033 eV Å$^{-1}$ (fig. S9c), which is much higher than that of Li-SP (0.014 eV Å$^{-1}$). These results indicate that Li-SP has more accurate results involved in surface configurations.

**S2. The Molecular Dynamic Simulations**

All of the molecular dynamic simulations were conducted using LAMMPS [17]. All configurations used in Li deposition were supercells of the primitive bcc cell with an experimental lattice constant of 3.51 Å [13, 16]. In the deposition simulation, eighteen layers of Li atoms are the substrate, with eight bottom layers of atoms fixed for a bulk environment. The NVT ensemble with the Langevin thermostat was used in all deposition simulations. A time step of 1 fs was used in all simulations. All visualizations of the MD trajectory were performed with the OVITO program [18], and the adaptive common neighbor analysis (a-CNA) method [19] was used to identify and distinguish a typical phase.

**Homogeneous deposition simulation conditions.** The size of the supercells (fig. S10) used in this simulation is 28.07 nm × 1.40 nm × 30.00 nm. Li atoms are generated at the top of the supercell, and the probability of atoms generation at X and Y coordinates obeys the distribution of X~U (0, 28.07) and Y~U (0, 1.40), where U stands for uniform distribution. The time interval of atom generation (generation rate) is 5 Li ps$^{-1}$, and the falling speed of lithium atoms is 500 m s$^{-1}$. These simulations were performed in a 300 K of NVT ensemble with a total deposition time of 3 ns.

**Inhomogeneous deposition simulation conditions.** The size of the supercells in inhomogeneous deposition is the same as the size in the homogeneous deposition (fig. S16, Fig.

2a, and Fig. 3). Lithium atoms were generated at the top of the supercell. The generation region was limited to two small regions in fig. S16 and Fig. 2a and one small region in Fig. 3. The generation rate and the falling rate are the same as homogeneous deposition conditions in Fig. 2a. The simulation of Fig. 2a was performed in a 100 K of NVT ensemble with a total deposition time of 3 ns. In Fig. 3, the falling speed of Li is proportional to the generation rate. For example, when the Gr is 0.5 Li ps$^{-1}$, the falling speed is 500 m s$^{-1}$. When 1 Li ps$^{-1}$, the falling speed is 1,000 m s$^{-1}$ to keep the distance between the two falling atoms.

**S3. Further investigated on homogeneous deposition.**

The temperature of the system was controlled at 100 K, 200 K, and 300 K, respectively, while other deposition conditions remain unchanged. The results show that the temperature has little effect on the final Li morphology in the homogeneous deposition (figs. S12 and S13), except that higher temperature generally results in a smoother surface (figs. S12c and S13c). We analyzed the local structure of Li atoms through the adaptive common neighbor analysis (a-CNA) method [19] and found that temperature would affect the degree of Li crystallization (fig. S13). At low temperature (~100 K), body-centered cubic (BCC) states occupy up to 92.7% of total states (fig. S10b), while at high temperature (~300 K), the amorphous state increases with the BCC state decreasing to 66.3% (fig. S13d). We also studied the influence of Gr on the morphology during deposition, which can be considered as current densities. The Gr are 1 Li ps$^{-1}$, 5 Li ps$^{-1}$, 20 Li ps$^{-1}$ with the deposition duration of 15 ns, 3 ns, 0.75 ns, respectively, to ensure the same number of deposited atoms. The results (fig. S14) show that the Gr has little effect on the morphology in the deposition, only the less BCC state at a high Gr (fig. S14c), related to the increased temperatures of the surface. Besides, the simulation with a larger 3D supercell also shows a surface self-healing process in the deposition, indicating the independence of the supercell shape (fig. S15).

**S4. Further discussion on inhomogeneous deposition**

**Discussion on coulombic interaction.** Coulombic interaction in the electric field was not considered in the simulation due to the limitation of the model. To overcome this deficiency, we simulate the dendrite growth process by limiting the deposition area, corresponding to the tip effect. Li ions in the electrolyte are affected by a relatively strong electric field in the actual deposition process. Once Li ions are deposited on the surface of lithium metal, Li ions will become Li atoms, and the influence of the electric field for Li atoms will be much weaker than that for the Li ions. Therefore, only considering the interaction between atoms can still give reliable simulation results [20]. Besides, we are focusing on studying the growth process of Li (dendrite) and the dynamic proprieties of Li atoms but not the falling process of Li ions. Therefore, our model can still give reliable simulation results without considering the effect of the electric field.

**Discussion on current density.** In the actual deposition process, the current density is much lower than this corresponding value of the generation rate (Gr, 1 Li ps$^{-1}$ at an area of 39.3×10$^{-14}$ cm$^2$ corresponding to ~4×10$^8$ mA cm$^{-2}$). Due to the limitation of simulation time, we cannot adjust the Gr (current density) to an order of magnitude close to the experimental value. Similar situations also exist in other Li deposition (growth) processes [20, 21]. Fortunately, we can still change the Gr within a certain range to study the influence on the dendrite shapes and deduce the dendrite change trend.

**Discussion on the surface temperature.** The higher surface temperature than the bulk comes from the condensation of deposited Li. The temperature may be overestimated than that of the actual deposition process, considering the heat exchange between the surface and electrolyte. In the actual deposition process, the energy released by every deposited Li is related to the overpotential. Generally, an overpotential of −0.5 V is able to reach a relatively high current density leading to additional energy of 0.5 eV for every Li atom [22]. Suppose all of the energy is converted to kinetic energy on the surface. In that case, the surface will have a temperature of 3,865 K without considering the exchange of heat, according to the thermodynamic formula: $E_k = 3/2\ kT$, where $E_k$ is the kinetic energy, $k$ is the Boltzmann constant, and $T$ is the temperature. However, heat exchange exists between the surface and the local environment (including the SEI, electrolyte, and lithium metal, et al.). Meanwhile, part of the overpotential energy is consumed in the electrolyte and the counter electrode. In our simulation, the heat exchange only existed between the dendrite and the substrate. Thus, when considering the heat exchange in the systems, this ultra-temperature less likely exists in the liquid electrolyte but more likely exists in the solid electrolyte systems, which is more similar to our simulation environment. The above discussion indicates that the surface temperature is usually higher than the bulk temperature in deposition and may reach the melting point at a high overpotential.

**Supplementary Figures**

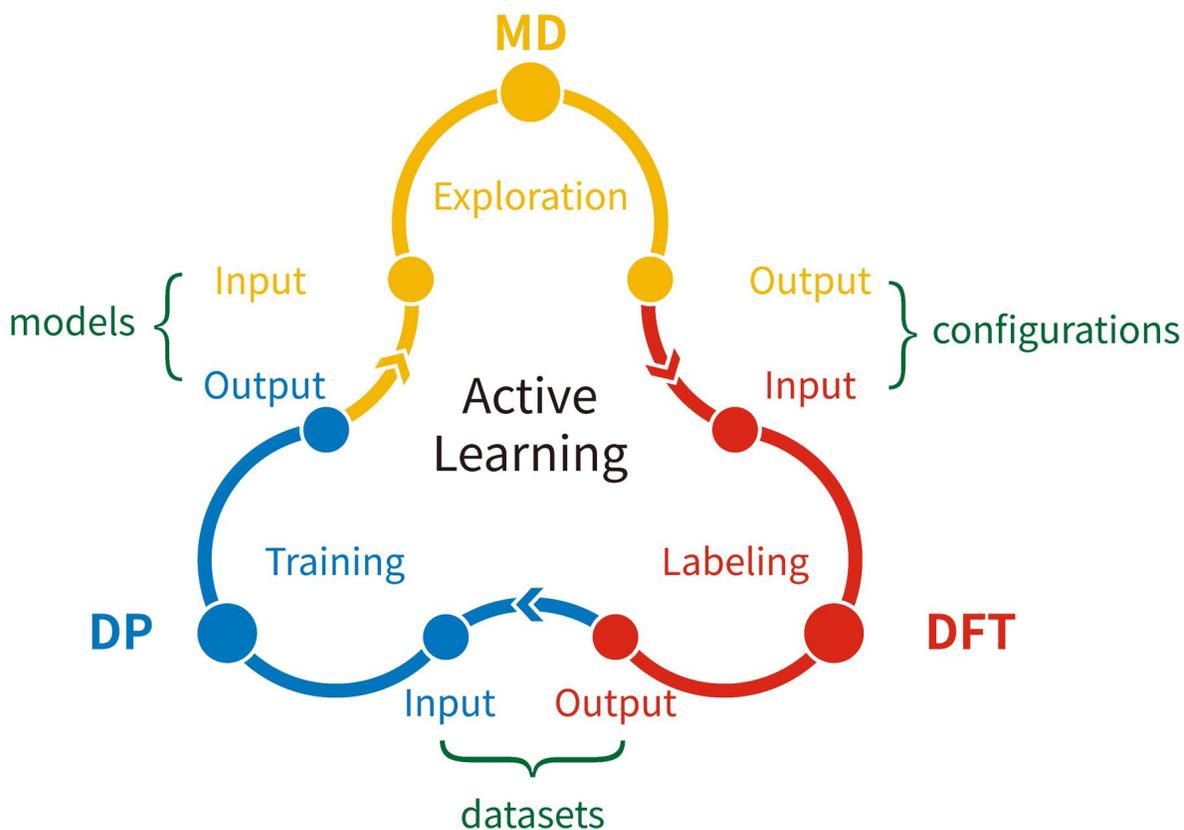

**Fig. S1. The schematic of active learning.**

The active learning contains three steps: exploration, labeling, and training. The DP stands for deep neural network potentials, MD stands for molecular dynamics simulation, and DFT stands for density functional theory calculations. These steps will repeat automatically until the accurate models are established.

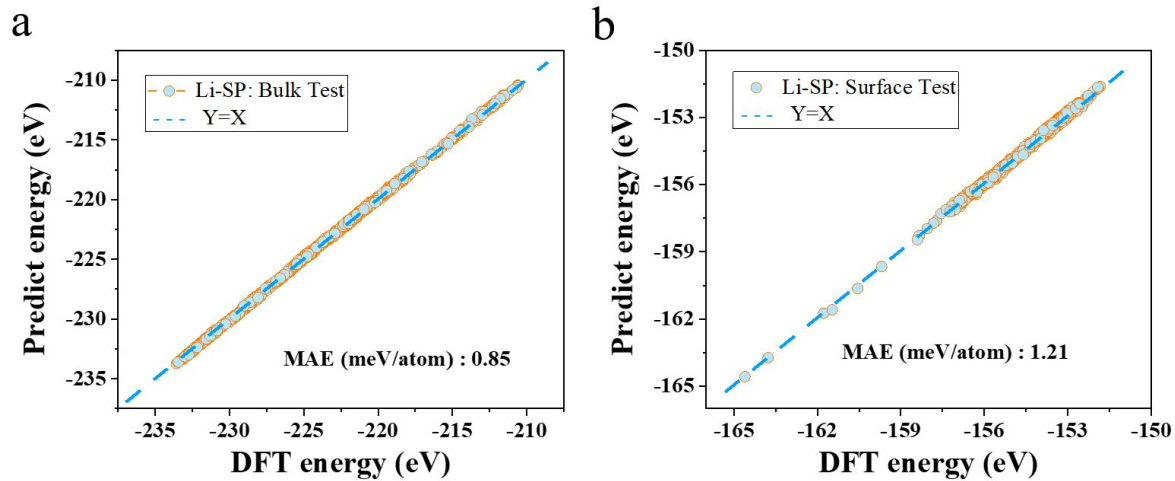

**Fig. S2. The energy results of the surface and bulk test datasets predicted by Li-SP and DFT calculation.**
The MAE is 0.85 meV/atom for bulk test (**a**) and 1.21 meV/atom for surface test (**b**).

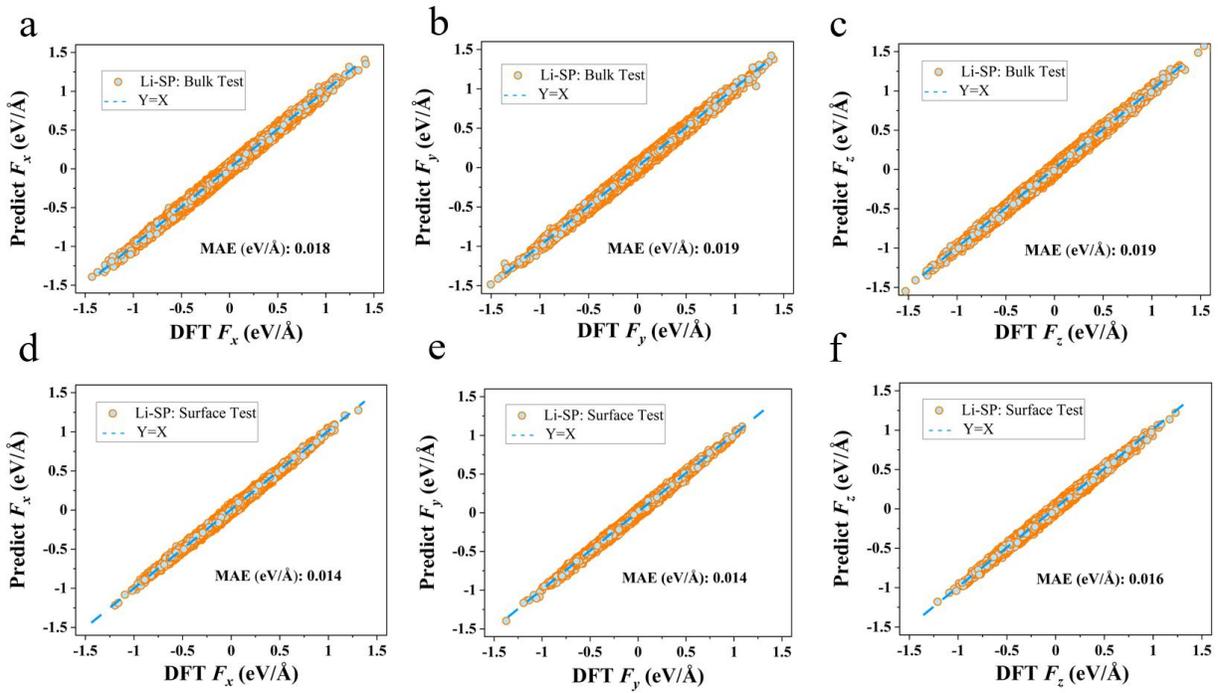

**Fig. S3. The force results of surface and bulk test dataset predicted by Li-SP and DFT calculation in x, y, and z direction.**

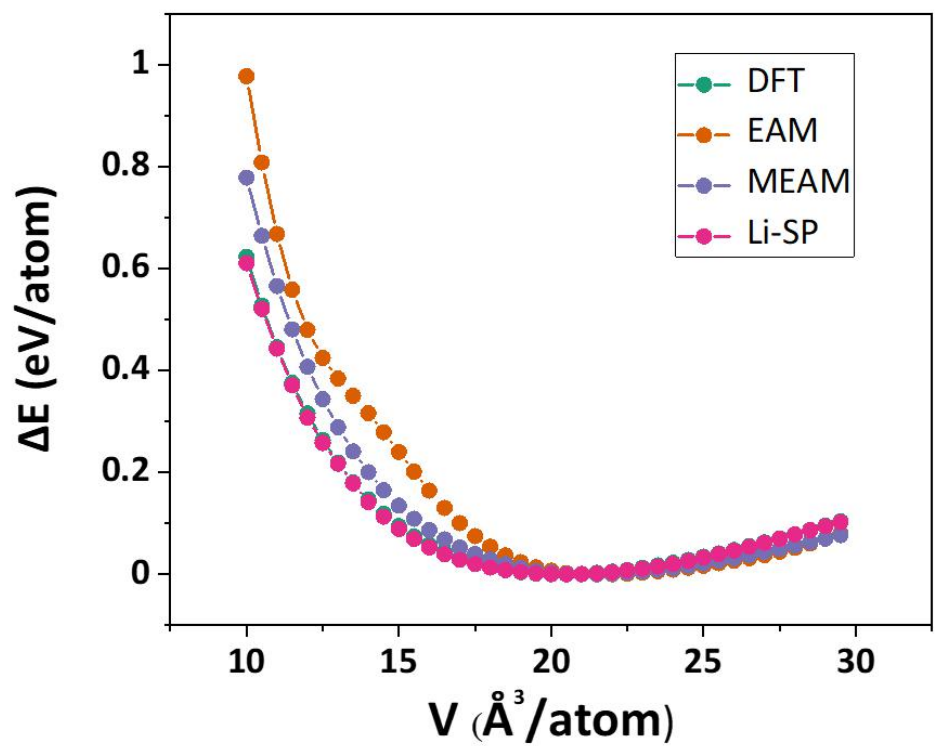

**Fig. S4. The Li Equation of state (EOS) obtained by different methods.**

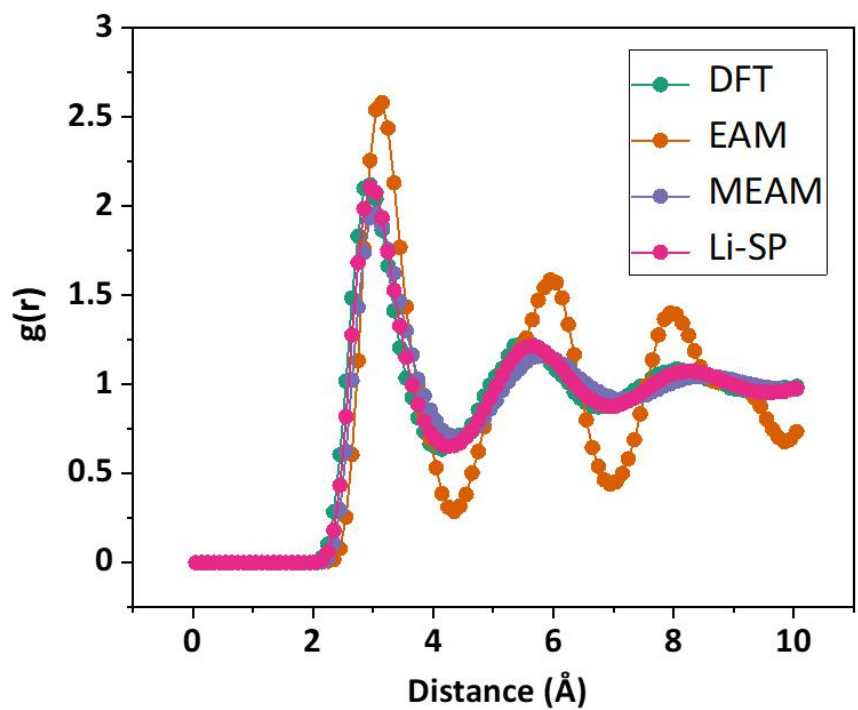

**Fig. S5. The radial distribution functions of Li measured at 700 K by the above methods.**

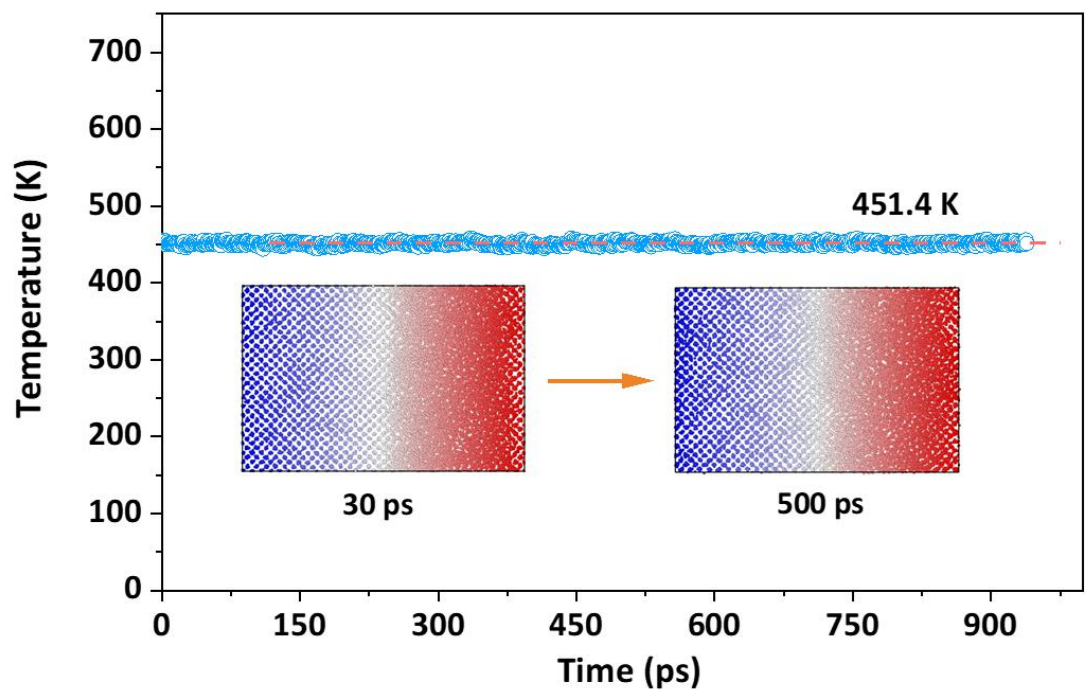

**Fig. S6. The melting point (451.6 K) of Li measured in solid-liquid mixed phase by Li-SP model.**

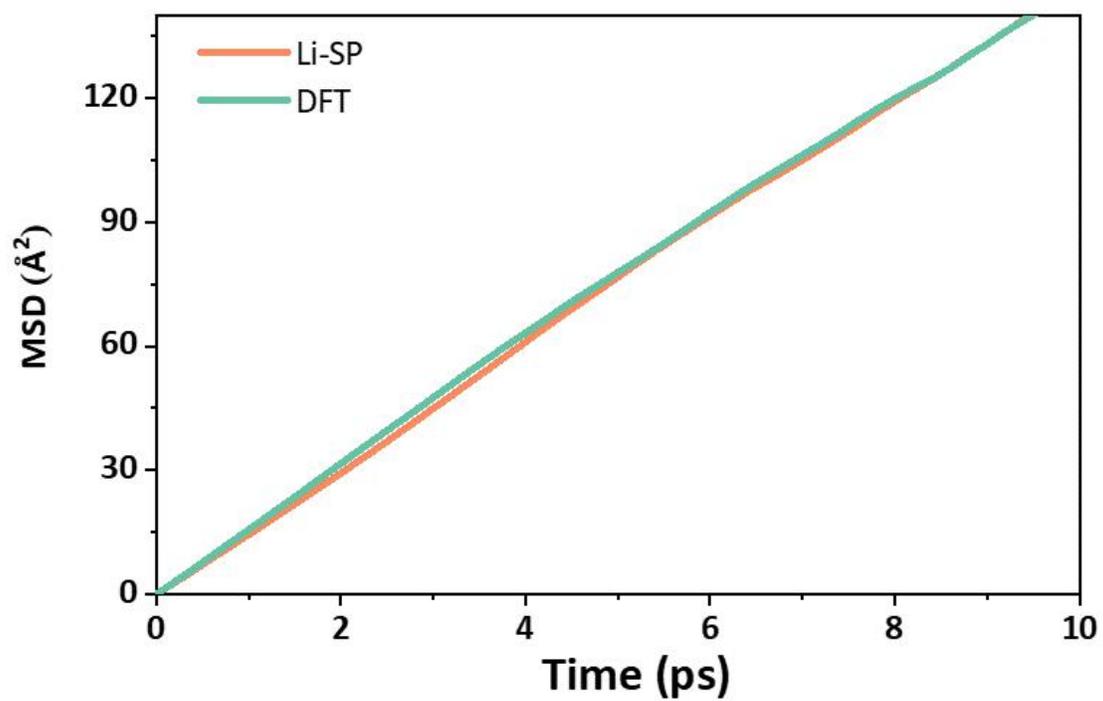

**Fig. S7. MSD calculated by Li-SP and DFT at 1,000 K.**

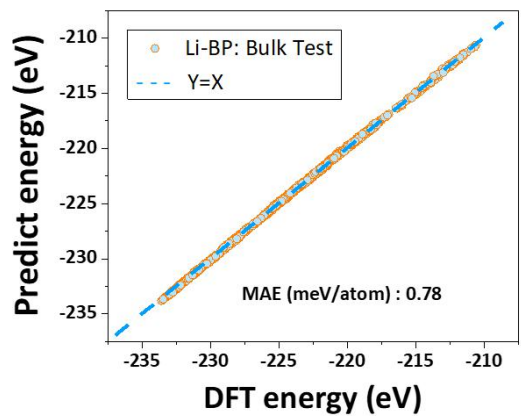 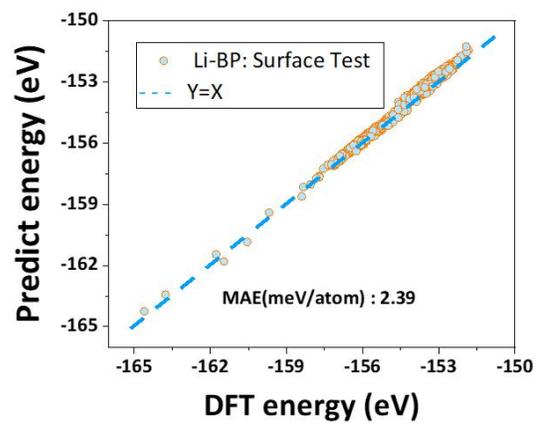

**Fig. S8. The energy test results predicted by lithium bulk potential (Li-BP) and DFT.** **(a)** the results of bulk test dataset; **(b)** the results of the surface test dataset.

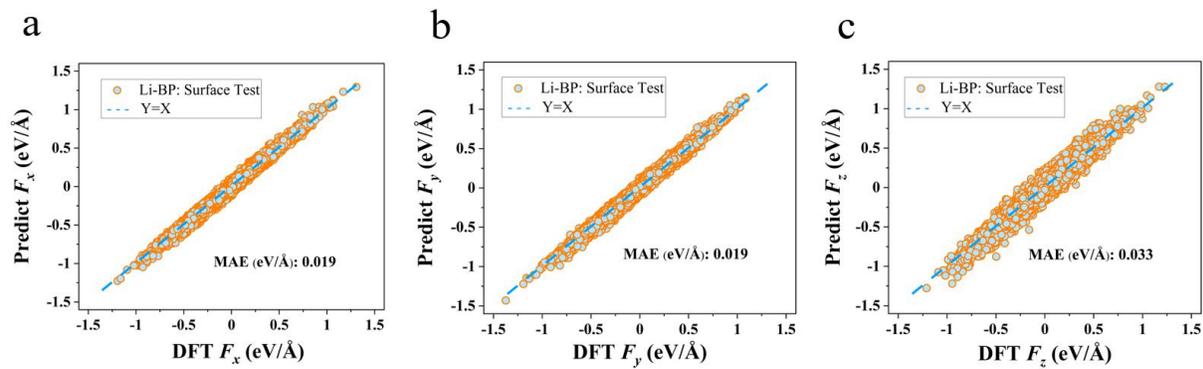

**Fig. S9. The force test of surface test dataset predicted by Li-DP and DFT calculation in three directions. (a) x, (b) y, and (c) z.**

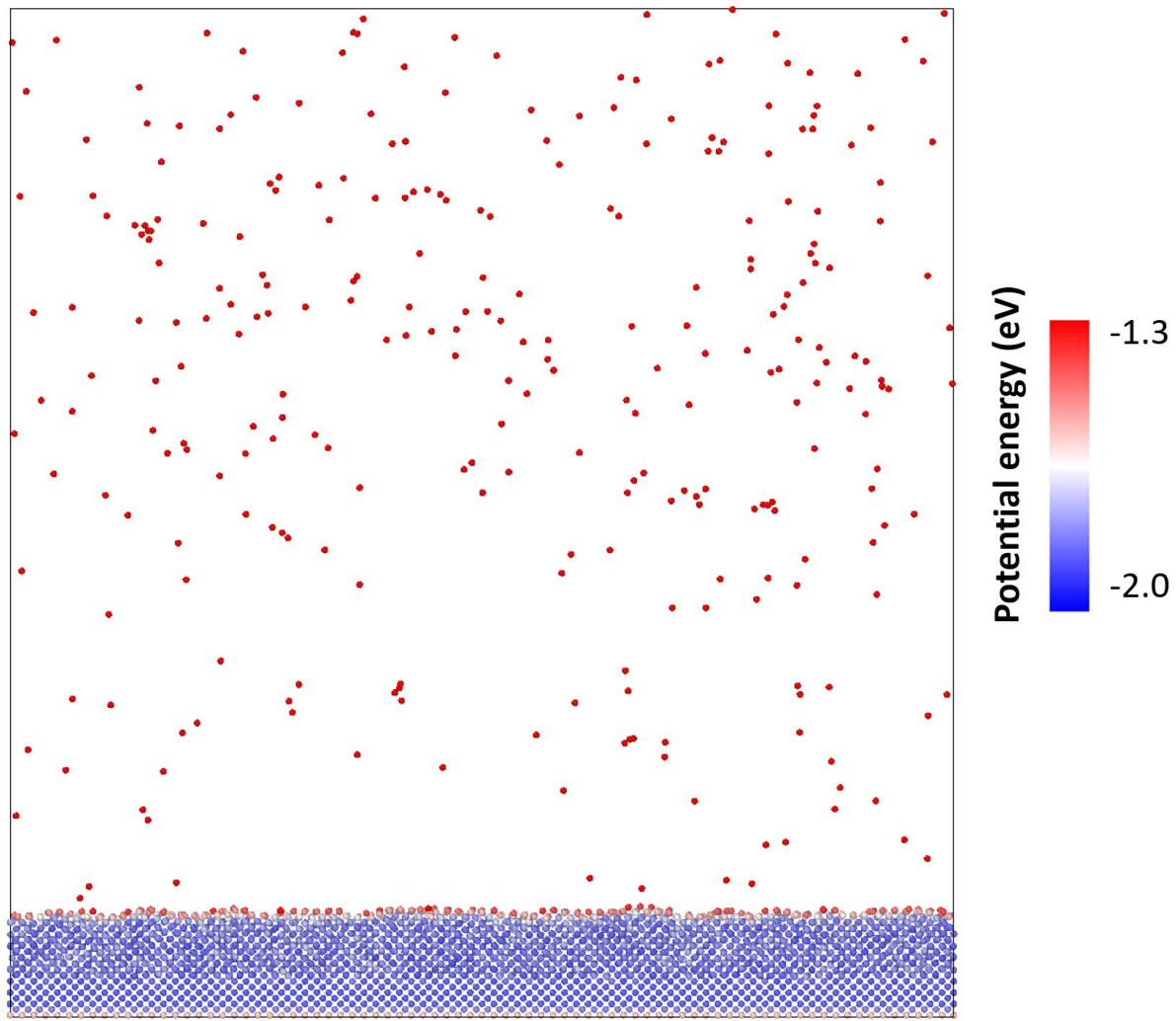

**Fig. S10. Schematic diagram of homogeneous deposition.**
Type or paste caption here. Create a page break and paste in the Figure above the caption.

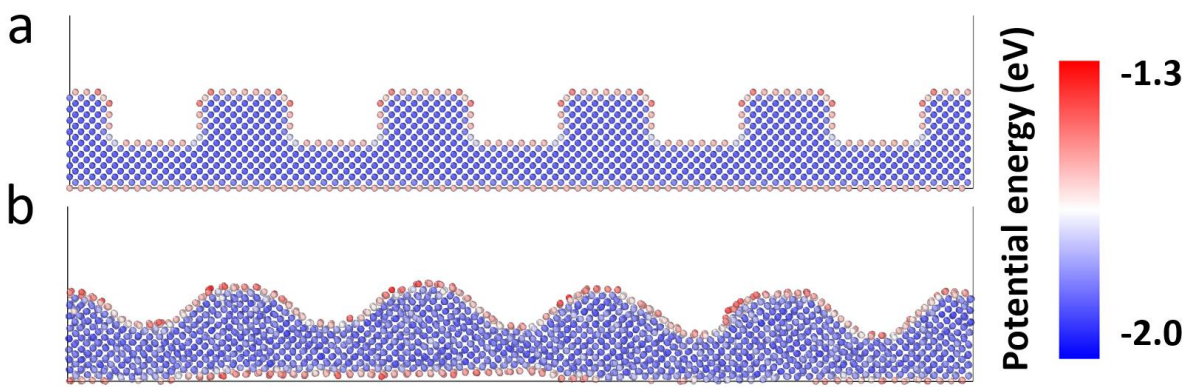

**Fig. S11. The evolution of the rectangular surface at 300 K.**
(**a**) the initial configuration, (**b**) the configuration with a stable corrugated surface after relaxation.

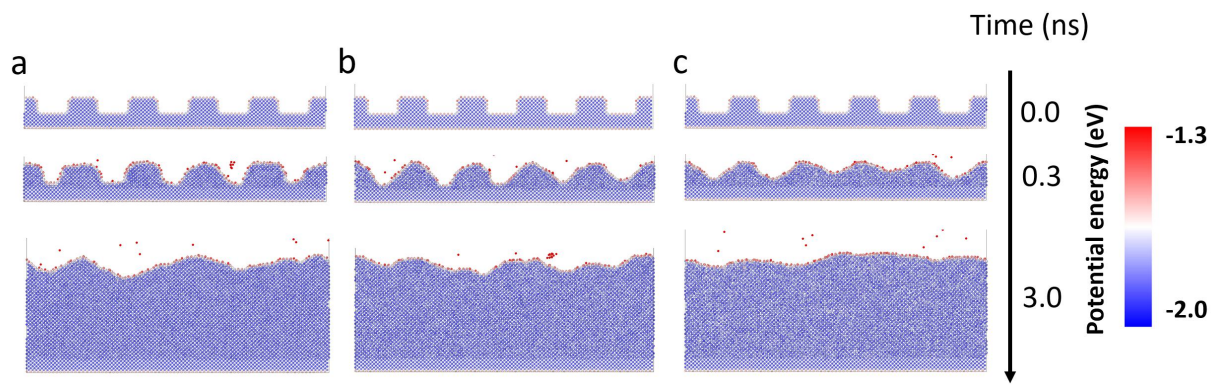

**Fig. S12. Snapshots of homogeneous deposition with a rectangular surface at different temperatures.**
(**a**) 100 K, (**b**) 200 K, and (**c**) 300 K in NVT ensemble. The generation rate is 5 Li ps$^{-1}$.

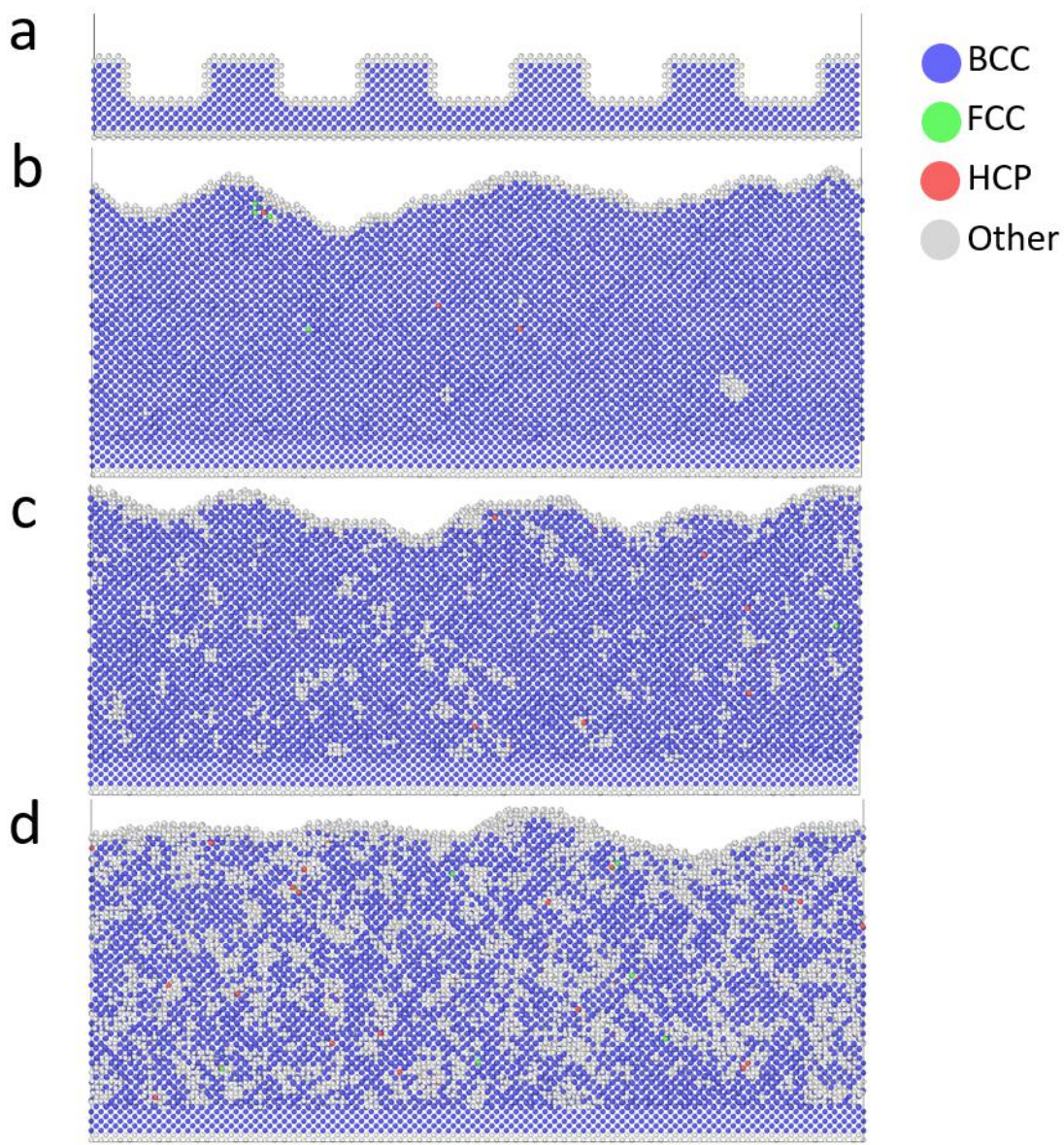

**Fig. S13. The Simulated results of homogeneous deposition at different temperatures by adaptive common neighbor analysis (a-CNA).**

(**a**) the initial configuration, (**b**) the snapshot at 3 ns and 100 K, (**c**) the snapshot at 3 ns and 200 K, and (**d**) the snapshot at 3 ns and 300 K. The blue, green, and red balls present body-centered cubic (BCC), face-centered cubic (FCC), and hexagonal close-packed (HCP), respectively. The gray ball presents other local environments, including surface and amorphous atoms.

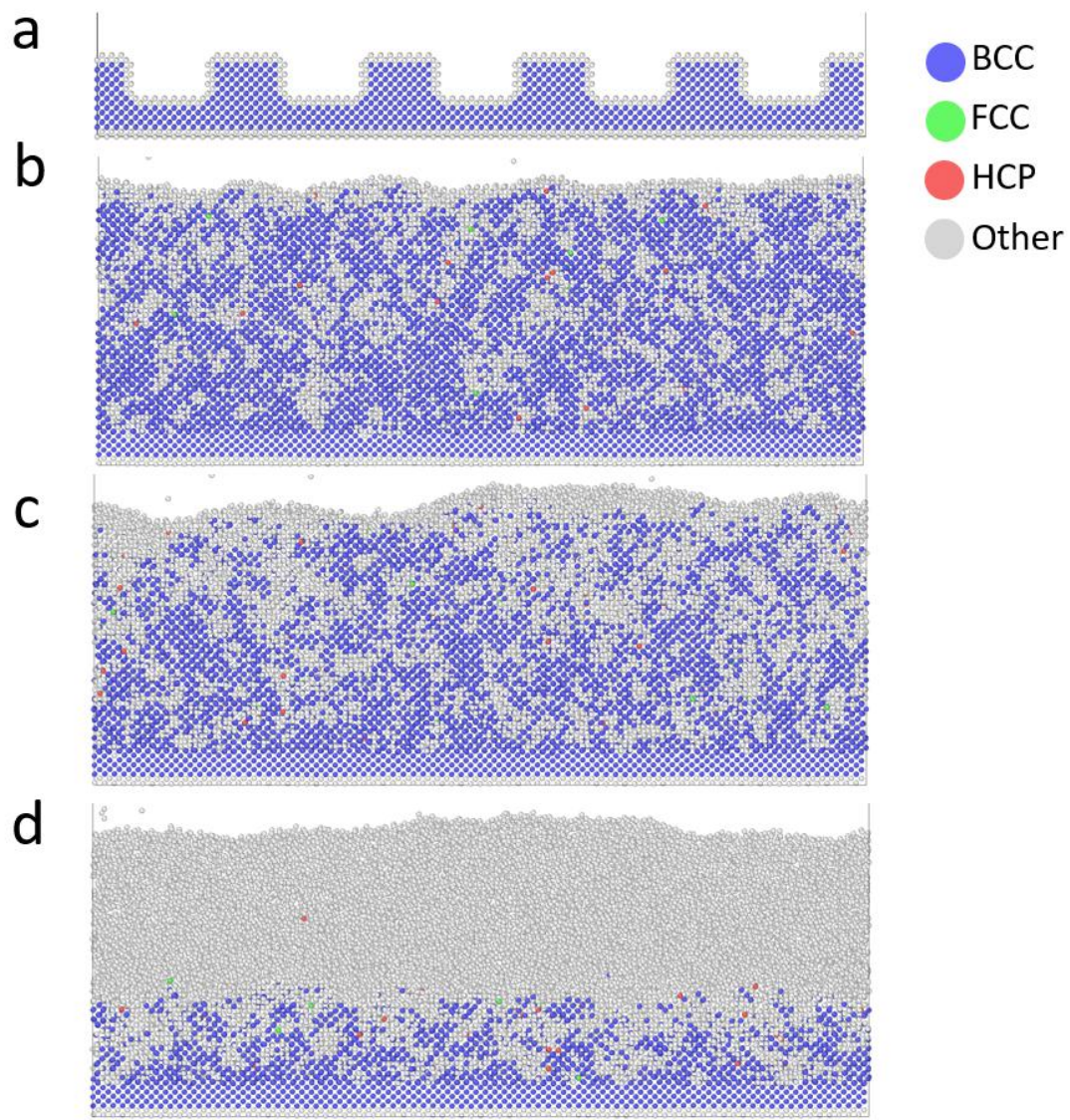

**Fig. S14. Simulated results of homogeneous deposition at different deposition rates.**
(**a**) 1 Li ps$^{-1}$, 15 ns (**b**) 5 Li ps$^{-1}$, 3 ns (**c**) 20 Li ps$^{-1}$, 0.75 ns. The number of deposition Li atoms is 15000. The surface of the initial structure is smooth, and the temperature is 300 K.

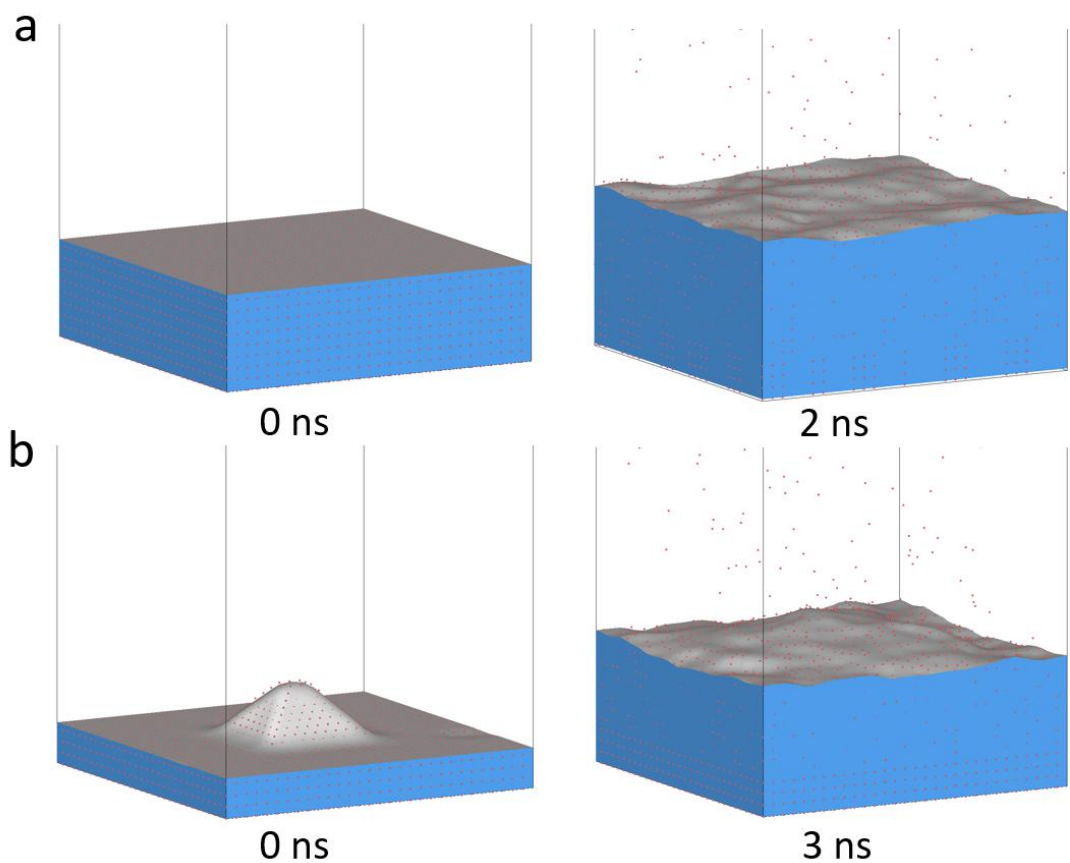

**Fig. S15. Simulated results of Li deposition under the configuration of a box of 10.5 nm (x) × 10.5 nm (y) × 30.0 nm (z) at 300 K, and the deposition rate was 5 Li ps$^{-1}$.**
(**a**) flat surface, (**b**) positive triangle surface.

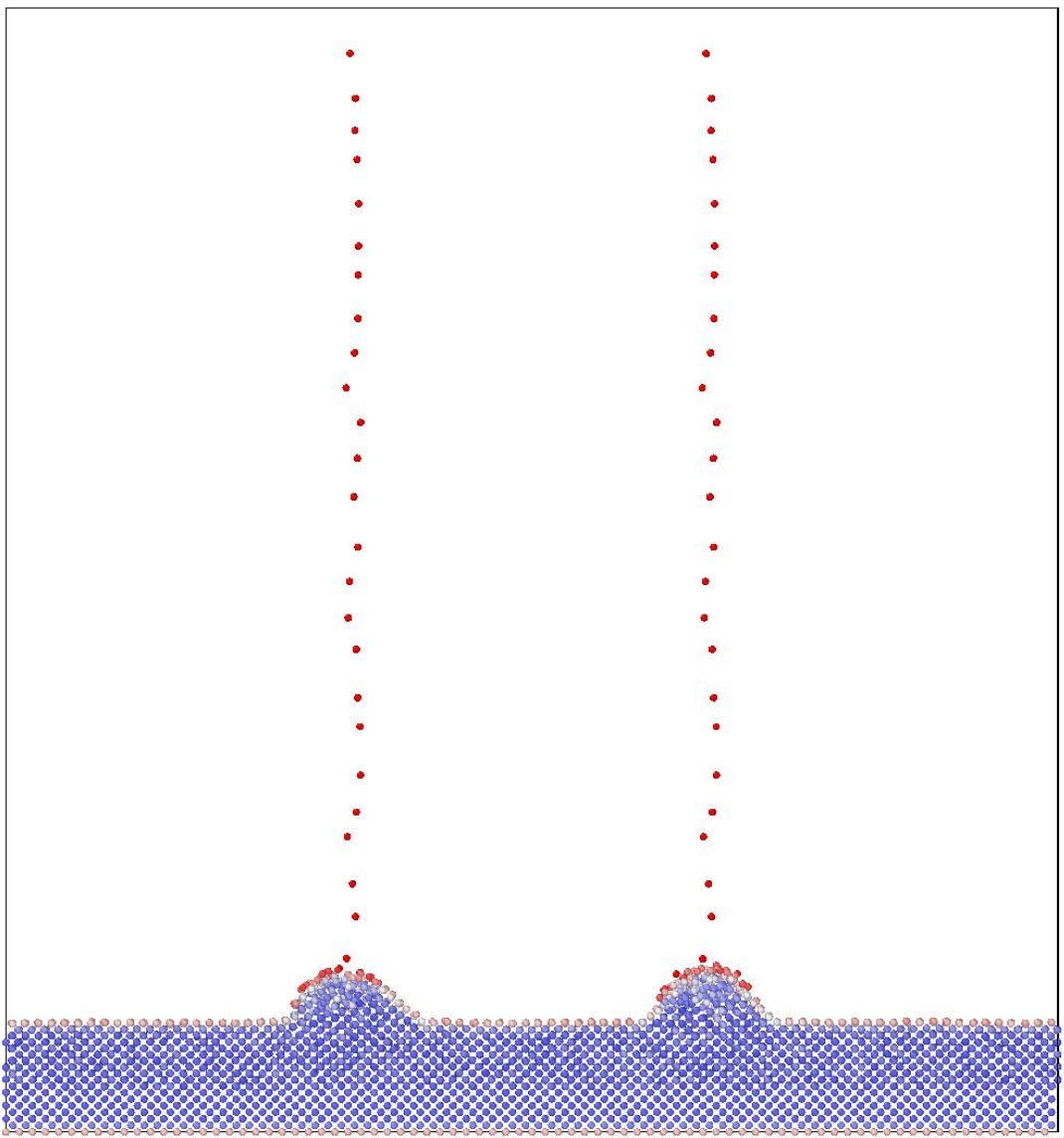

**Fig. S16. Schematic diagram of inhomogeneous deposition.**

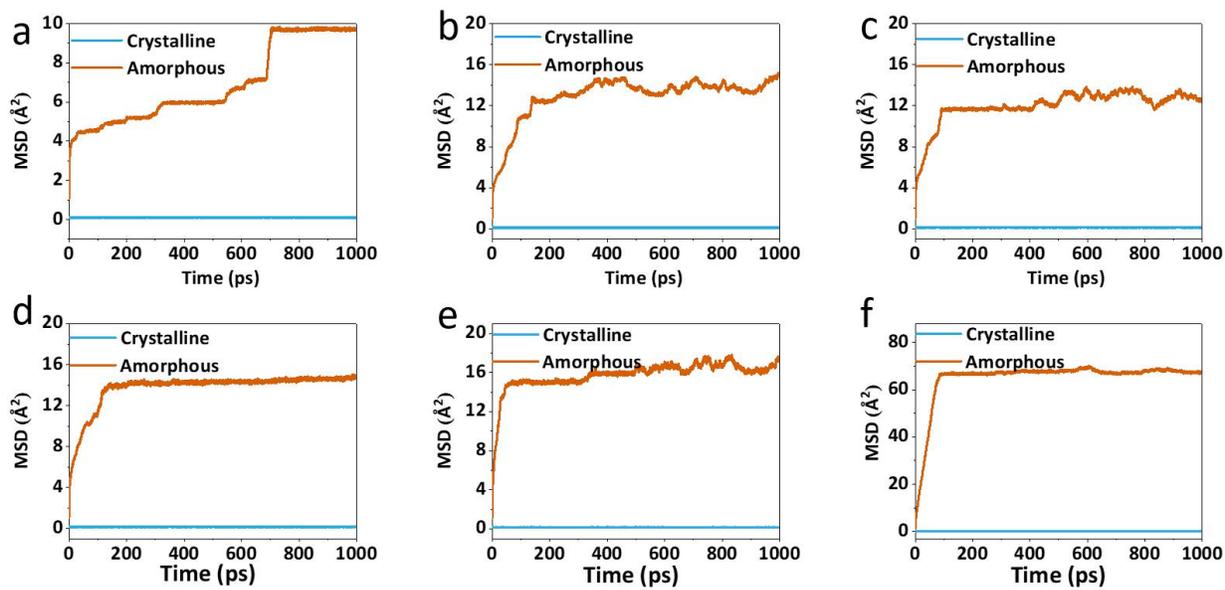

**Fig. S17. Mean square displacement comparison (MSD) of amorphous state and crystalline state (initial structure) at different temperatures.**
(**a**) 50 K; (**b**) 100 K; (**c**) 150 K; (**d**) 200 K; (**e**) 250 K; (**f**) 300 K. The initial crystalline configuration contains 2,000 Li atoms with a 10×10×10 of pristine BCC structure. The initial amorphous configuration was obtained by MD of the initial crystalline configuration at 1,000 K.

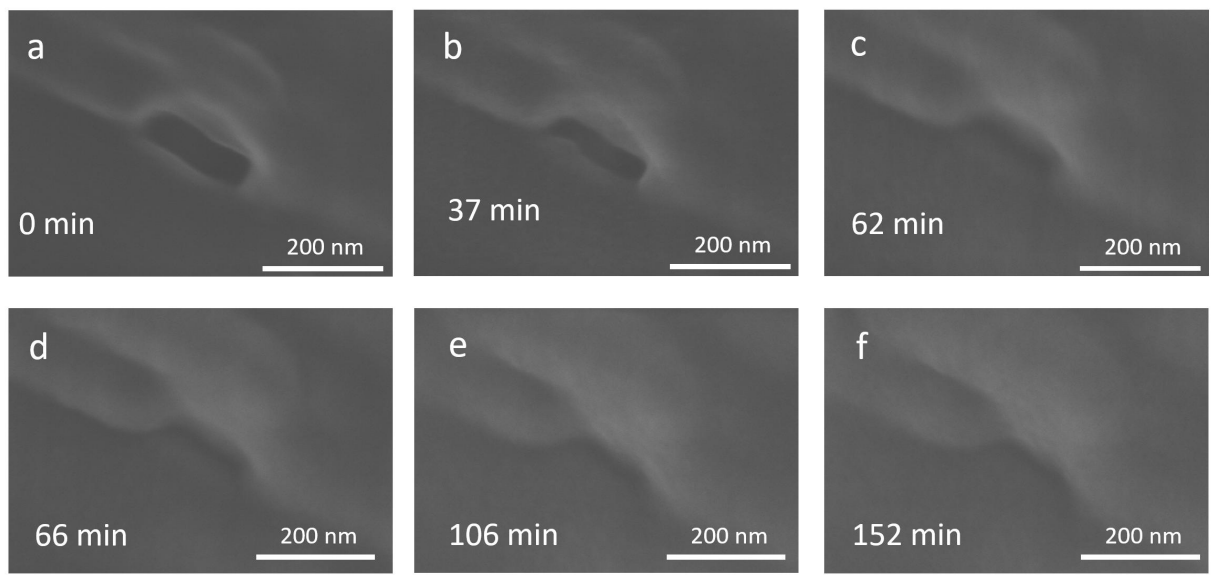

**Fig. S18. Li bulk self-healing process under the scanning electron microscope.**

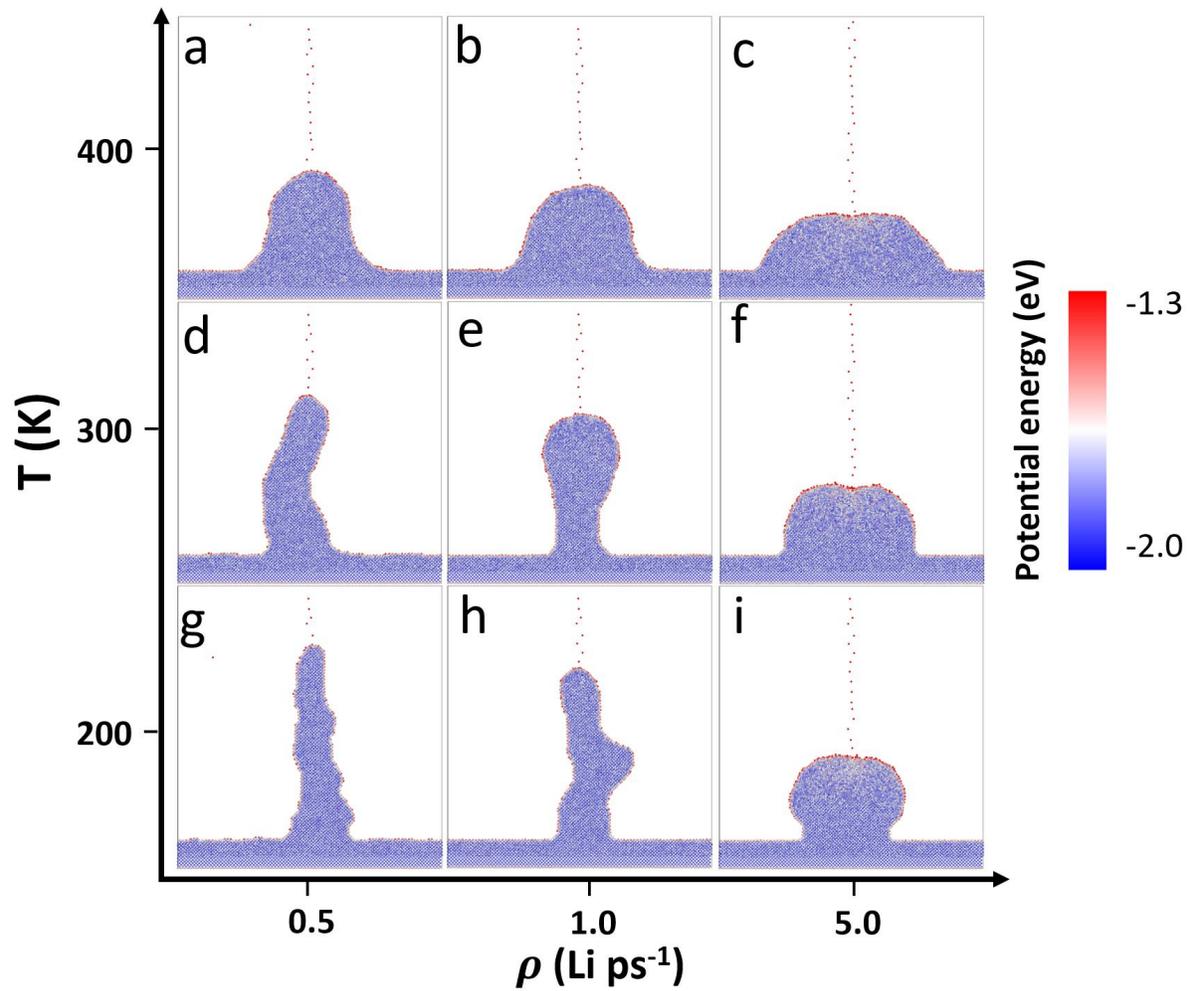

**Fig. S19.** The snapshots of the inhomogeneous deposition with ~6000 deposited Li atoms in different temperatures and Gr.

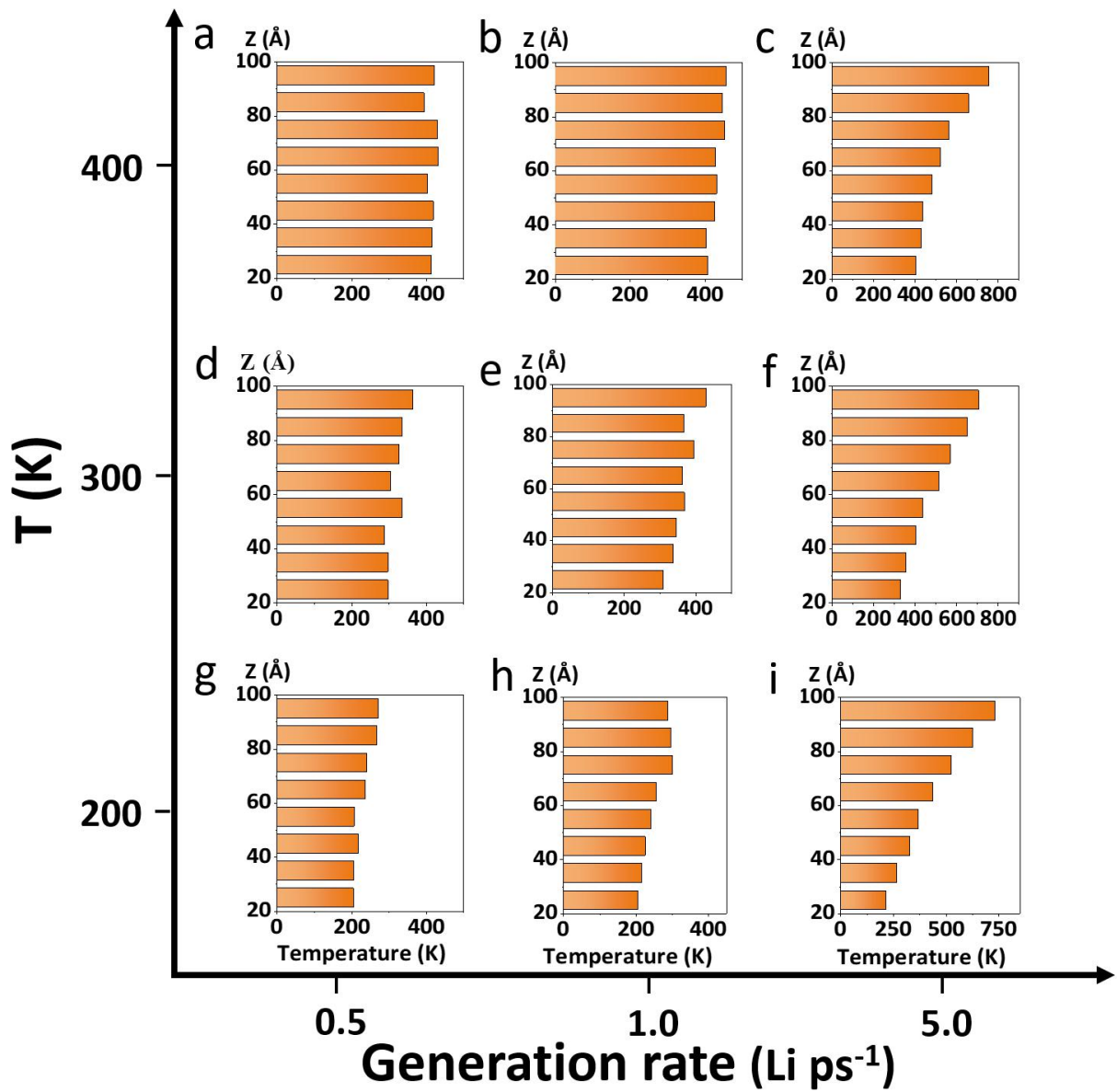

**Fig. S20. The temperature at different dendrite altitudes corresponding to Fig. 3.**
The local temperatures of the dendrite were calculated by dividing the dendrite into several sub-regions according to the size of the Z-axis. The divided interval of the Z-axis is 1 nm.

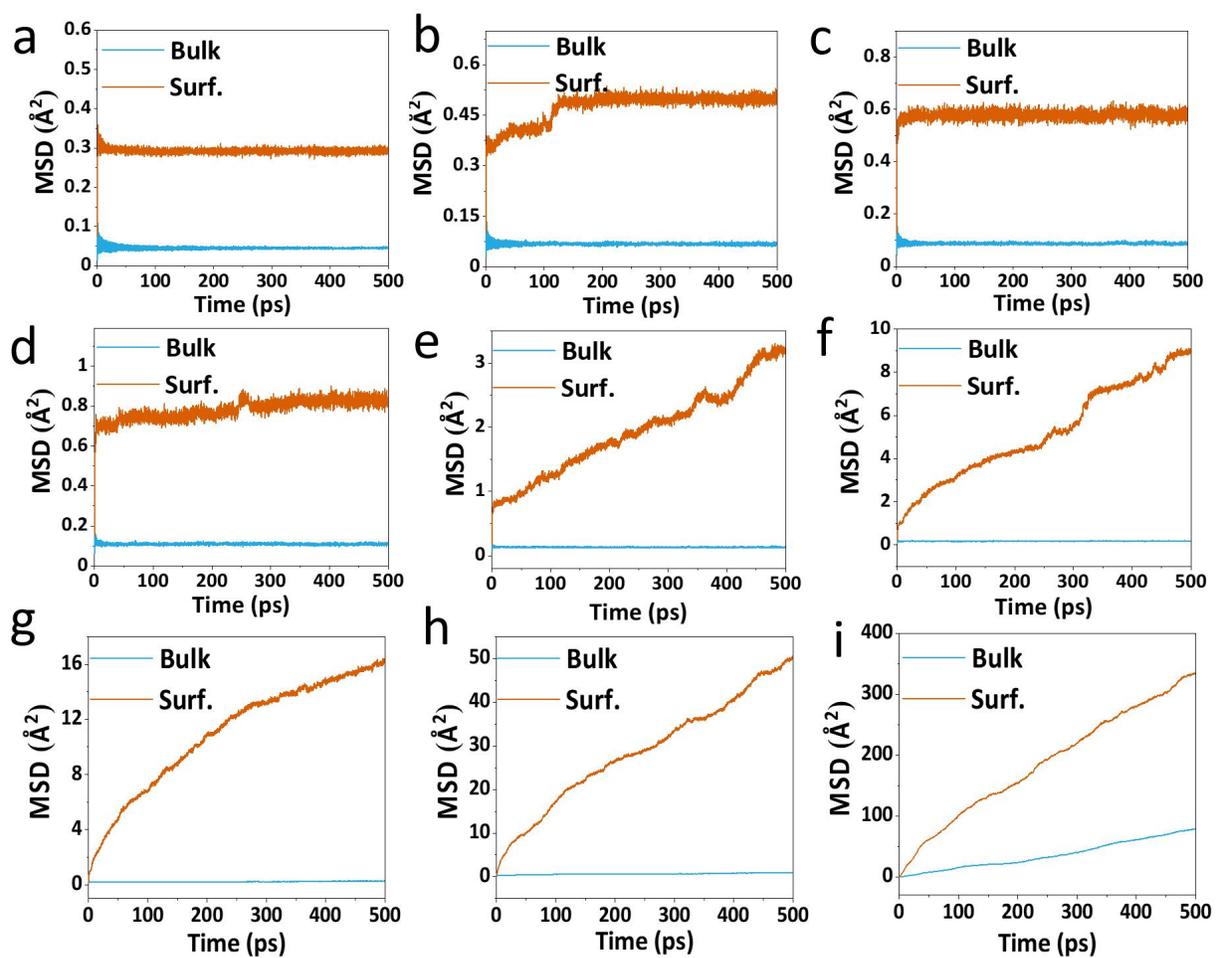

**Fig. S21. MSD for bulk and surface atoms at different temperatures.**
(**a**) 50 K, (**b**) 100 K, (**c**) 150 K, (**d**) 200 K, (**e**) 250 K, (**f**) 300 K, (**f**) 350 K, (**h**) 400 K, and (**i**) 450 K.

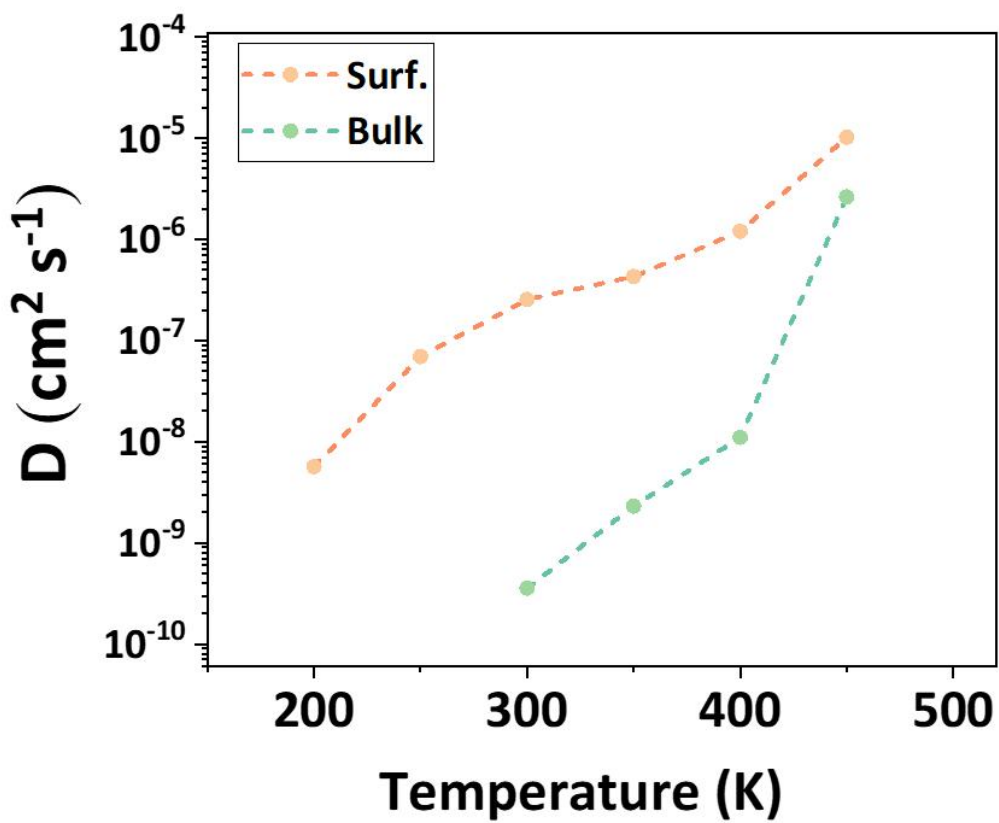

**Fig. S22.** The diffusion coefficient of surface and bulk Li atoms with different temperatures.

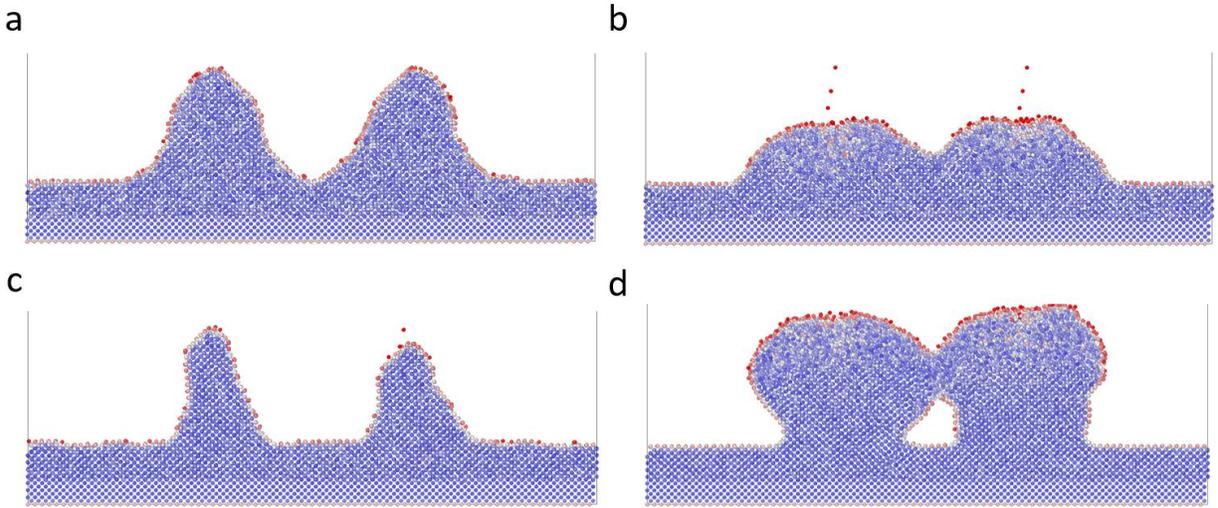

**Fig. S23. The schematic diagram of bulk self-healing in different temperatures and Gr.** (**a**) 350 K at 1 Li ps$^{-1}$; (**b**) 300 K at 5 Li ps$^{-1}$; (**c**) 200 K at 1 Li ps$^{-1}$; and (**d**) 100 K at 5 Li ps$^{-1}$.

# Supplementary Tables

**Table S1. Physical properties of Li predicted by various models.**

The physical properties include elastic constants ($C_{ij}$), bulk modulus ($B_v$), shear modulus ($G_v$), surface energy with three low Miller indexes, and the melting point ($T_m$). The test models are DFT, EAM, MEAM, Li-SP, Li-BP, and experiments. The Li-SP did not show obvious advantages when compared with EAM and MEAN in the calculation of some elastic constants for the reason that our training datasets have none information about elastic constants and modulus, while both EAM and MEAN take elastic constants and modulus as the targets to fit the potential parameters [13, 14].

|  |  | DFT | EAM (g) | MEAM (h) | Li-SP | Li-BP | Exp. |
|---|---|---|---|---|---|---|---|
| Elastic (GPa) | $C_{11}$ | 15 (a) | 14.74 | 14.96 | 16.36 | 19.45 | 14.8 (b) |
|  | $C_{12}$ | 13 (a) | 12.49 | 12.45 | 11.72 | 14.46 | 12.5 (b) |
|  | $C_{44}$ | 11 (a) | 10.73 | 10.33 | 10.87 | 12.02 | 10.8 (b) |
| Modulus (GPa) | $B_V$ | 14 (a) | 12.57 | 13.29 | 13.27 | 16.12 | / |
|  | $G_V$ | 7 (a) | 6.89 | 6.70 | 7.45 | 8.21 | / |
| Surface Energy (J/m²) | 111 | 0.54 (c) | 0.460 | 0.473 | 0.545 | 0.575 | / |
|  | 110 | 0.50 (c) | 0.354 | 0.404 | 0.481 | 0.487 | / |
|  | 100 | 0.46 (c) | 0.361 | 0.397 | 0.463 | 0.503 | / |
| $T_m$ (K) |  | / | 660 (d) | 450 ± 10 (e) | 451.6 | 445 | 454 (f) |

a. *ref.* [23]
b. *ref.* [14, 16]
c. *ref.* [24]
d. *ref.* [13]
e. *ref.* [14]
f. *ref.* [15]

# Supplementary References

1. Kresse G, Furthmüller J. Efficient iterative schemes for ab initio total-energy calculations using a plane-wave basis set. *Phys. Rev. B* 1996, **54**(16)**:** 11169-11186.

2. Kresse G, Furthmüller J. Efficiency of ab-initio total energy calculations for metals and semiconductors using a plane-wave basis set. *Comput. Mater. Sci.* 1996, **6**(1)**:** 15-50.

3. Perdew JP, Burke K, Ernzerhof M. Generalized Gradient Approximation Made Simple. *Phys. Rev. Lett.* 1996, **77**(18)**:** 3865-3868.

4. Blöchl PE. Projector augmented-wave method. *Phys. Rev. B* 1994, **50**(24)**:** 17953-17979.

5. Zhang L, Han J, Wang H, Car R, E W. Deep Potential Molecular Dynamics: A Scalable Model with the Accuracy of Quantum Mechanics. *Phys. Rev. Lett.* 2018, **120**(14)**:** 143001.

6. Wang H, Zhang L, Han J, E W. DeePMD-kit: A deep learning package for many-body potential energy representation and molecular dynamics. *Comput. Phys. Commun.* 2018, **228:** 178-184.

7. Zhang L, Lin D-Y, Wang H, Car R, E W. Active learning of uniformly accurate interatomic potentials for materials simulation. *Phys. Rev. Mater.* 2019, **3**(2)**:** 023804.

8. Zhang Y, Wang H, Chen W, Zeng J, Zhang L, Wang H*, et al.* DP-GEN: A concurrent learning platform for the generation of reliable deep learning based potential energy models. *Comput. Phys. Commun.* 2020, **253:** 107206.

9. Zhang LF, Han JQ, Wang H, Saidi WA, Car R, E WN. End-to-end Symmetry Preserving Inter-atomic Potential Energy Model for Finite and Extended Systems. In: Bengio S, Wallach H, Larochelle H, Grauman K, CesaBianchi N, Garnett R (eds). *Advances in Neural Information Processing Systems 31*, vol. 31. Neural Information Processing Systems (Nips): La Jolla, 2018.

10. He KM, Zhang XY, Ren SQ, Sun J. Deep Residual Learning for Image Recognition. *Proceedings of the IEEE Conference on Computer Vision and Pattern Recognition (CVPR)* 2016**:** 770-778.

11. Kingma DP, Ba J. Adam: A method for stochastic optimization. *arXiv preprint arXiv:1412.6980* 2014.

12. Wu J, Zhang Y, Zhang L, Liu S. Deep learning of accurate force field of ferroelectric HfO2. *Phys. Rev. B* 2021, **103**(2)**:** 024108.

13. Nichol A, Ackland GJ. Property trends in simple metals: An empirical potential approach. *Phys. Rev. B* 2016, **93**(18)**:** 184101.

14. Cui Z, Gao F, Cui Z, Qu J. Developing a second nearest-neighbor modified embedded atom method interatomic potential for lithium. *Modell. Simul. Mater. Sci. Eng.* 2011, **20**(1)**:** 015014.

15. Brandes EA, Brook G, Eds. *Smithells Metals Reference Book*. Butterworth-Heinemann, Oxford, ed. 7, 1992.

16. Ko W-S, Jeon JB. Interatomic potential that describes martensitic phase transformations in pure lithium. *Comput. Mater. Sci.* 2017, **129:** 202-210.

17. Plimpton S. Fast Parallel Algorithms for Short-Range Molecular Dynamics. *J. Comput. Phys.* 1995, **117**(1)**:** 1-19.

18. Stukowski A. Visualization and analysis of atomistic simulation data with OVITO-the Open Visualization Tool. *Modell. Simul. Mater. Sci. Eng.* 2010, **18**(1)**:** 015012.

19. Stukowski A. Structure identification methods for atomistic simulations of crystalline materials. *Modell. Simul. Mater. Sci. Eng.* 2012, **20**(4)**:** 045021.

20. Yang M, Liu Y, Nolan AM, Mo Y. Interfacial Atomistic Mechanisms of Lithium Metal Stripping and Plating in Solid-State Batteries. *Adv. Mater.* 2021, **33**(11)**:** 2008081.

21. Wang X, Pawar G, Li Y, Ren X, Zhang M, Lu B, *et al.* Glassy Li metal anode for high-performance rechargeable Li batteries. *Nat. Mater.* 2020, **19**(12)**:** 1339-1345.

22. Li L, Basu S, Wang Y, Chen Z, Hundekar P, Wang B, *et al.* Self-heating–induced healing of lithium dendrites. *Science* 2018, **359**(6383)**:** 1513-1516.

23. de Jong M, Chen W, Angsten T, Jain A, Notestine R, Gamst A, *et al.* Charting the complete elastic properties of inorganic crystalline compounds. *Sci. Data* 2015, **2**.

24. Tran R, Xu ZH, Radhakrishnan B, Winston D, Sun WH, Persson KA, *et al.* Surface energies of elemental crystals. *Sci. Data* 2016, **3**.